\documentclass[fleqn,usenatbib]{mnras}

\usepackage{newtxtext,newtxmath}

\usepackage[T1]{fontenc}

\DeclareRobustCommand{\VAN}[3]{#2}
\let\VANthebibliography\thebibliography
\def\thebibliography{\DeclareRobustCommand{\VAN}[3]{##3}\VANthebibliography}

\usepackage{graphicx}	
\usepackage{amsmath}	

\title[blazar induced pair beams]{The role of resonant plasma instabilities in the evolution of blazar induced pair beams}

\author[Perry et al.]{
Roy Perry$^{1}$, Yuri Lyubarsky$^{1}$
\\
$^{1}$Physics Department, Ben-Gurion University, P.O.B. 653, Beer-Sheva 84105, Israel}

\date{Accepted XXX. Received YYY; in original form ZZZ}

\pubyear{2021}

\newcommand{\eqb}{\begin{equation}}
\newcommand{\eqe}{\end{equation}}
\begin{document}
\label{firstpage}
\pagerange{\pageref{firstpage}--\pageref{lastpage}}
\maketitle

\begin{abstract}
The fate of relativistic pair beams produced in the intergalactic medium by very-high energy emission from blazars remains controversial in the literature. The possible role of resonance beam plasma instability has been studied both analytically and numerically but no consensus has been reached. In this paper, we thoroughly analyze the development of this type of instability. This analysis takes into account that a highly relativistic beam loses energy only due to interactions with the plasma waves propagating within the opening angle of the beam (we call them parallel waves), whereas excitation of oblique waves results merely in an angular spreading of the beam, which reduces the instability growth rate.  For parallel waves, the growth rate is a few times larger than for oblique ones, so they grow faster than oblique waves and drain energy from the beam before it expands. However, the specific property of extragalactic beams is that they are extraordinarily narrow; the opening angle is only $\Delta\theta\sim 10^{-6}-10^{-5}$. In this case, the width of the resonance for parallel waves, $\propto\Delta\theta^2$, is too small for them to grow in realistic conditions. We perform both analytical estimates and numerical simulations in the quasilinear regime. These show that for extragalactic beams, the growth of the waves is incapable of taking a significant portion of the beam's energy. This type of instability could at best lead to an expansion of the beam by some factor but the beam's energy remains nearly intact.
\end{abstract}


\section{Introduction}

The high-energy relativistic jets from blazars emit gamma ray radiation, which travels along in the same direction as the jets. 
During their course through the intergalactic medium (IGM), the very high energy (VHE) gamma photons interact with the extragalactic background light (EBL) and the cosmic microwave background (CMB). This interaction has a mean free path ranging from several Mpc to 1 Gpc, which leads to the production of ultra-relativistic electron-positron pairs
with Lorentz factors of about $\gamma\sim 10^4 - 10^7$. These relativistic beams propagate in the IGM and are expected to lose energy by inverse-Compton scattering upon the CMB. 
However, the observed gamma-ray bump associated with the product photons of inverse-Compton cascades (ICC) is lower than predicted (e.g. \citealt{Aharonian2006, Neronov2010, Broderick12}). One possible explanation is pair beam deflection by the inter-galactic magnetic field. Measurement of the GeV flux on telescopes in conjunction with the estimation of blazar-originated gamma rays flux, a lower bound on the intergalactic magnetic field is predicted. The inconsistencies between the predicted and measured flux are attributed to the deflection of particles by the magnetic field \citep{Archambault_2017,Tiede_2017,Ackermann_2018}. An alternate possibility \citep{Broderick12, Schlickeiser2012} is that before the beam pairs up-scatter photons, they lose energy through excitation of plasma waves via 
the resonance beam-plasma instability.
The beam-plasma instability results in excitation of plasma oscillations (Langmuir waves) with a frequency  close to the plasma frequency (e.g., \citealt{Krall1973}). 
The IGM density is of the order of $10^{-7} \text{cm}^{-3}$, so the plasma frequency is of the order of a few Hz, while the beam density ranges over many orders of magnitude, though estimated to be no more than $10^{-21} \text{cm}^{-3}$ (e.g., \citealt{Broderick12}). 
The instability could develop in two different regimes: 
reactive (hydrodynamic) and kinetic (e.g., \citealt{Boyd2003}). The first applies to an essentially cold beam, whose thermal spread is small enough such that all the particles are in resonance with the wave. The second corresponds to the case of a wide spread in the distribution function of the beam. In that case, only a fraction of the particles contribute to the growth of the wave.
If the beam is initially narrow enough such that the instability develops in the hydrodynamic regime, then mostly oblique waves are excited.
Therefore the beam expands while losing almost no energy \citep{Fainberg1970}. Eventually, the momentum spread of the beam becomes large enough 
for the kinetic regime to come into play. 
During the kinetic regime, the beam could lose most of its energy by exciting plasma waves which eventually heats the background plasma
\citep{Fainberg1970,Rudakov70,Breizman71}.
The literature on the collective plasma processes in extragalactic pair beams is extensive but its conclusions still remain controversial. \citet{Miniati2013} studied different mechanisms that affect the development of resonance instability and concluded that for typical beam parameters, either non-linear wave interactions or inhomogeneities of the background plasma suppress this instability. 
However, no consensus has been reached about the role of the collective plasma processes in the evolution of extragalactic beams (e.g., \citealt{Broderick12, Broderick2014, Schlickeiser2012, Schlickeiser2013, Miniati2013, Chang2014,Krakau2014, Supsar2014, Chang2016, Sironi2014,Kempf2016,Rafighi2017, Broderick2018, shalaby2018growth, Vafin2018, Vafin2019, Yan2018, Batista2019, Shalaby2020}). 

The results of PIC simulations could not be interpreted unambiguously because the width of the wave-particle resonance is so small in extragalactic beams that it could not be resolved in simulations. Moreover, the numerical noise in simulations is high so the beam parameters had to be chosen such that the level of the excited plasma waves were high enough. In this case different non-linear processes play a significant role, so the results could hardly be rescaled to the realistic range of parameters. 

Up until now, the discussion was focused on the conditions for the development of beam instability. It has been taken for granted that if the instability develops in the kinetic regime, it is capable of decelerating the beam. In this paper, we take into account a subtle fact that only waves propagating within the open angle of the beam are responsible for energy loss in the beam whereas excitation of oblique waves merely results in an increase in its angular spread, which reduces the instability rate. Our main goal is to study the pair beam's evolution and behaviour in time, paying special attention to the relevant range of parameters for intergalactic beams. Namely, an extremely small growth rate and an extremely small angular opening of the beam. On the one hand, the last is sufficient for a transition to the kinetic instability regime (due to the very small growth rate), but on the other hand, it makes the width of the wave particle resonance extremely small, which is crucial in the evolution of the instability. 

The instability of relativistic beams was comprehensively studied in the early plasma literature \citep{Bludman1960_2,Bludman1960_1,Fainberg1970,Breizman71, Rudakov70}. However, most of the attention was given to relatively powerful beams, for which the kinetic regime is established only when the beam angular width is not too small. In these cases the beam efficiently loses energy in the kinetic regime. For weak intergalactic beams, the instability develops in the kinetic regime when the beams are extremely narrow, in which case the conditions for energy transfer from the beam to the plasma waves become unrealistically restrictive. We show that even if the instability develops, the beam's energy remains practically unchanged, and its angular speard expands just a few times over.

This paper is organized as follows: In section 2 we introduce the beam-plasma instability mechanism. 
In section 3 we describe the hydrodynamic and kinetic regimes of the instability and give explicit formulas for the growth rates. 
In section 4 we estimate the conditions for the instability and for the effective energy loss of the beam for realistic IGM parameters. In section 5 we present numerical simulations of the beam evolution by solving the kinetic equations in the quasi-linear regime. In section 6 we conclude and summarize our work.

\section{Preliminary Considerations}
\subsection{Beam-wave resonance}{\label{sec:resonance}}

The beam-plasma instability (e.g., \citealt{Krall1973}) results in excitation of plasma oscillations (Langmuir waves)
satisfying the resonance (Cerenkov) condition:
\begin{equation}
 \omega  = {\bf{k}} \cdot {\bf{v}},
\label{resonance}\end{equation}
where $\omega$  and ${\bf{k}}$ are the frequency and wave vector of the wave. 
For dilute beams, $n_b \ll n_p$, the frequency of the excited waves is close to the plasma frequency,
\begin{equation}
\omega_{p} = \sqrt{\frac{4\pi n_p e ^2}{m_e}},
\end{equation}
where $m_e$, $e$ are the electron mass and charge, respectively, and $n_p$ is the plasma density.

The extragalactic beams are highly energetic; the typical particle Lorentz factor is $\gamma\sim 10^4-10^7$. Even though the beam energy spread is large,
$\Delta\gamma\sim\gamma$, the velocity spread is negligibly small, $\Delta v\sim\Delta\gamma/\gamma^3$. Therefore 
 we can describe each beam particle velocity in spherical coordinates (on the azimuthal plane $\phi = 0$) as
\begin{equation}
{\bf{v}} = c\left(1-\theta^2/2,\theta\right),
\end{equation}
where the first and the second components correspond to the particle velocity components along  and perpendicular to the beam direction, respectively.
The 
wave-vector, ${\bf{k}}$, can be described in a three-dimensional spherical coordinates space as:
\begin{equation}
{\bf{k}} = \left(k,\theta_0,\phi_0 \right).
\end{equation}
The resonance condition (\ref{resonance}) 
can then be written as:
\begin{equation}
\cos\phi_0 = \frac{\omega_{p}/kc - \cos\theta \cos\theta_0 }{\sin\theta \sin\theta_0}.
\end{equation}
One can therefore describe the resonance wave-vector $\bf{k}$ with two independent parameters, $k$ and $\theta_0$.

\begin{figure}
\centering
\includegraphics[width=\columnwidth]{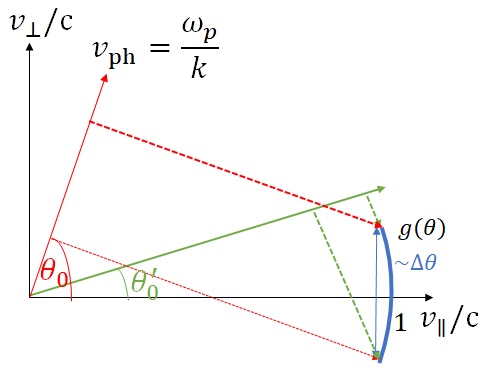}
\caption{(Color online)
The beam-wave resonance condition in velocity space. Shown are the particle velocity distribution (thick blue arc with an opening angle $\Delta\theta\ll 1$) and two plasma waves at two angles $\theta_0\gg \Delta\theta$ and $\theta'_0\sim\Delta\theta$ (red and green arrows). The range of the phase velocities which are in resonance with the beam is larger for the first wave than for the second one.}
\label{fig:RESONANCE}
\end{figure}

The geometry of the  particle-wave resonance in velocity space is shown in Figure \ref{fig:RESONANCE}. 
The arrows represent the directions of propagation of the Langmuir waves, where $\theta_0, \theta'_0$ are the angles of these vectors with respect to the beam axis, chosen as the horizontal axis. The arc represents the distribution of the pair-beam particles, ranging over an angular spread of $\Delta\theta \ll 1$. 
The range of resonant waves is signified with the bordered arrow between the two dashed lines. 
For any $\theta_0$, only waves in this range of $\bf{k}$ can exchange energy and momentum with the beam particles. The figure shows that the range of resonant waves, $\Delta k$, decreases with decreasing wave vector angle. Indeed, to first order in $\Delta \theta$, the range of resonance $\Delta k$ behaves as $\sim \tan\theta_0 \Delta\theta $. Therefore $\Delta k/k\sim \Delta\theta$ at $\theta_0\sim 1$, but when the wave propagates within the opening angle of the beam, $\theta_0\sim\Delta\theta$, the resonance becomes very narrow, $\Delta k/k\sim \Delta\theta^2$.

The angular spread of intergalactic beams is very small. After pair-production, the angular distribution of the particles is quite narrow because of their direct emission from a point-like source, far away from the intergalactic plasma.
Initially, the spread is about $\Delta\theta \sim \gamma^{-1}$. The beam could expand as a result of the instability but not by much. Therefore the width of the resonance is very small even for oblique waves, whereas for waves nearly aligned with the beam, it is extraordinarily small. We show below that this has a profound effect on the evolution of the beam because on the one hand, only nearly aligned waves may be responsible for any energy loss of the beam, but on the other hand, 
they easily violate the resonance conditions. Therefore, their growth may be easily suppressed by weak perturbations, such as inhomogeneity of the background plasma or/and non-linear wave interactions. 

\subsection{Energy loss vs expansion; qualitative consideration}
The particle distribution of the beam evolves because of the induced emission and absorption of plasma waves. The energy and momentum conservation principles in a single emission/absorption event are described via the following equations:
 \begin{equation}
     E=E'\pm\hbar\omega_p;\qquad {\bf{p}}={\bf{p}}'\pm\hbar{\bf{k}}.
 \end{equation}
Here we denote the parameters of the beam particle before and after the wave excitation by unprimed and primed quantities, respectively. For concreteness, consider an emission event and assume for simplicity that the beam axis, the particle velocity and the wave vector are all in the same plane. This imples:
\begin{equation}
\delta\gamma = \gamma' - \gamma = -\frac{\hbar\omega_p}{mc^2},
\end{equation}
\begin{equation}{\label{eq:MOM_CON}}
\begin{array}{l}
 m\gamma 'v'\sin \theta ' - m\gamma v\sin \theta  =  - \hbar k\sin {\theta _0} ,\\ 
 m\gamma 'v'\cos \theta ' - m\gamma v\cos \theta  =  - \hbar k\cos {\theta _0} ,\\ 
 \end{array}
\end{equation}
where $\theta$, $\theta_0$ are the particle and wave vector angles with respect to the beam axis, respectively. 
We consider a very narrow,
$\theta,\theta' \ll 1$, highly relativistic, $\gamma\gg 1$, beam  and expand up to first order. 
The pair of momentum equations now become:
\begin{equation}{\label{eq:MOM_CON2}}
\begin{aligned}
 &\cos \left( {\theta  - {\theta _0}} \right)\left( {\gamma '\theta ' - \gamma \theta } \right) =  - \frac{{\hbar {\omega _p}}}{{m{c^2}}}\sin {\theta _0} ,\\ 
 &\cos \left( {\theta  - {\theta _0}} \right)\left( {\gamma ' - \gamma } \right) =  - \frac{{\hbar {\omega _p}}}{{m{c^2}}}\cos {\theta _0} ,
 \end{aligned}
 \end{equation}
Upon dividing the first equation in (\ref{eq:MOM_CON2}) by the second one and rearranging terms, we get the ratio between relative angular and energetic change,
\begin{equation}{\label{eq:REL_CHANGE}}
\left| {\frac{{\delta \theta }}{\theta }\frac{\gamma }{{\delta \gamma }}} \right| = \frac{{\tan {\theta _0}}}{\theta }.
\end{equation}
Eq. (\ref{eq:REL_CHANGE}) shows that the outcome of the emission/absorption process strongly depends on the relative angles between the particle and the wave. Namely, the relative change in energy is comparable to the relative change in angle only when the wave is directed within the opening angle of the beam, $\theta_0\sim \Delta\theta$. In the case $\theta_0\gg\Delta\theta$, the particle's angle varies significantly more than its energy. 

This conclusion may be considered a preliminary one because it is based only on analysis of a single emission/absorption event. The particle distribution function evolves as a result of many interactions. The evolution may be described as diffusion in the momentum space (e.g., \citealt{BreizmanREVIEW}).  In section \ 5.1, we show rigorously, by analysing the momentum diffusion coefficients, that only quasi-parallel waves ($\theta_0\sim\Delta\theta$) take away any energy efficiently whereas the interaction with oblique waves ($\theta_0\gg\Delta\theta$) leads only to beam expansion.

The evolution of the beam is governed by the total spectrum of resonant plasma oscillations. The dominant part of the spectrum will determine whether the beam loses energy or expands in angle. The instability mechanism will enhance the spectrum of waves in each direction according to their respective growth rate. Those which grow the fastest are the ones who primarily control the evolution of the beam.
Since the growth is exponential, even a moderate excess in growth rate will make the corresponding range of the wave spectrum dominant. It follows from eq. (\ref{eq:REL_CHANGE}), that the beam will lose its energy,
if small angle waves ($\theta_0 \sim \Delta\theta$) grow faster. Indeed, according to kinetic linear theory, the growth rate is larger for such waves (\citealt{Breizman71,Rudakov70}, see also section \ref{sec:KIN}).
However, the intergalactic beams are very narrow. Therefore the resonance width is especially narrow, $\sim\Delta\theta^2$, for these waves, as we have seen in the previous subsection. In this case, even a weak inhomogeneity 
of the IGM or/and weak nonlinear processes result in loss of the resonance between 
these waves and the beam, which suppresses their growth rate. 

\subsection{Effect of plasma inhomogeneity}{\label{sec:inhomogeneity}}

When a wave propagates through a steady but weakly inhomogeneous plasma, 
the wave vector varies according to the geometrical optics prescription
 \begin{equation}
     \frac{d\mathbf{k} }{dt}=-
     \nabla\omega_p.
 \end{equation}
 Taking into account that $k\sim\omega_p/c$ one sees that
if the characteristic scale of the density variation is $L$, such that $\vert\nabla\omega_p\vert=\omega_p/L$, the relative change in the wave vector after time $t$ would be

\begin{equation}
    \frac{\delta k}{k}\sim\frac{ct}L.
\label{k-drift}\end{equation}
Let a wave be excited initially in the resonance range, bounded by the two dashed lines in Figure \ref{fig:RESONANCE}. As it travels through the plasma, its wave number moves along that line, towards one of the resonance boundaries, until it crosses that boundary and exits the resonance region. If the exit time is smaller than the instability time, the instability does not develop \citep{Breizman70}. 
\citet{Miniati2013} have shown that inhomogeneity and other effects (non-linearity, collisional damping) 
impose strong constraints on the development of the instability of intergalactic beams. Since then, these effects were studied by many authors quoted in our introduction. In this paper, we show that the constraints on the energy exchange between the beam and the background plasma are much more severe because of a special role played by quasi-parallel waves. We have seen 
that the resonance width decreases as the angle between the wave and beam directions decreases. 
Hence, the inhomogeneity causes a loss of resonance much faster for quasi-parallel waves than for oblique waves. 
We concentrate on the effect of inhomogeneity. 
Non-linear effects could also suppress the instability by removing the waves from the resonance. However, 
 we show in section \ref{sec:IGM} that in the context of intergalactic beams, inhomogeneity alone completely suppresses the generation of quasiparallel waves, responsible for energy loss of the beam.  

It was claimed recently \citep{shalaby2018growth,Shalaby2020} that the inhomogeneity does not suppress the instability.  The authors considered a monochromatic 1D beam propagating through the 
 density distribution with a local minimum and applied periodic boundary conditions. 
They claim that within a "bowl", a standing plasma wave is formed between two turning points, 
which remain in resonance with the beam all the time. 
This artificial example could not be considered a disproof of the general effect outlined above.
First of all, the instability develops only in a small region in the vicinity of the local density minimum point. In the the 3D case, the effects of wave locking are irrelevant for most of the system's volume, even if there are many density "bowls". Moreover, in 1D simulations with periodic boundary conditions, the beam enters the bowl already appropriately modulated during the previous passages, therefore the growth of perturbations proceeds further, which does not happen in the real 3D world.

Additionally, it is worth stressing that the work by \citealt{shalaby2018growth,Shalaby2020} is based on an implicit assumption  that standing waves are formed in the region of interest. In 1D, standing waves are formed by back and forward directed waves bouncing between the turning points. In 3D, the wave changes direction at the turning point and therefore could not remain in resonance with the beam. The group velocity of such a non-resonant plasma wave is extremely small so that 
the global standing wave, in which the phases and the wave vectors of local oscillations are adjusted within the whole region, 
are not formed in the system.
In any case, even if the theory by \citet{shalaby2018growth,Shalaby2020} is universally correct, their instability growth rate is still smaller than the collisional decay rate of the plasma waves in the IGM. In sect.\ 4 below, where parameters of the beam are discussed,  we present the corresponding estimate.

\section{Beam-plasma instability}
The growth rate of the instability $\Gamma$ can be calculated by making use of Maxwell's equations and the linearized Vlasov equation (e.g., \citealt{Krall1973}). 
It is common in the literature to describe this instability mechanism in two different  regimes, i.e. the hydro (reactive) regime and the kinetic regime. 
\subsection{Hydrodynamic regime}
The hydrodynamic regime applies to waves which resonate with the beam as a whole. 
In other words, the beam is narrow enough such that the parallel velocity component of all the particles matches the phase velocity of the wave.  For every wave vector whose parallel component satisfies $v_{b} k_{\parallel} = \omega$, where $v_{b}$ is the typical velocity of the beam particles, the interaction is unstable but waves with different perpendicular components grow at different rates. 

The dispersion relation for unstable Langmuir waves in the electrostatic approximation looks like \citep{Bludman1960_2}:
\begin{equation}
1 - \frac{{\omega _p^2}}{{{\omega ^2}}} - \frac{{\omega _b^2}}{{{{\left( {\omega  - {k_\parallel }{{\rm{v}}_b}} \right)}^2}}}\left( {\frac{{{\gamma ^2}{{\sin }^2}{\theta _0} + {{\cos }^2}{\theta _0}}}{{{\gamma ^3}}}} \right) = 0,
\end{equation}
where $\omega_{b}=\sqrt{4\pi e^2 n_b/m_e}$ is the beam plasma frequency.
 There is a wide range of unstable solutions of this equation but in the case of interest, $n_b/n_p\sim 10^{-15}-10^{-18}$, $\gamma\sim 10^6$,
only the plasma waves in resonance with the beam grow fast enough  (see Appendix A). 
The maximal growth rate of the resonant plasma wave, is \citep{Bludman1960_2,Fainberg1970}
\begin{equation}{\label{eq:GR_HYDRO}}
{\Gamma _{{\rm{hydro}}}} = \frac{{\sqrt 3 }}{{{2^{4/3}}}}\frac{{{\omega _p}}}{\gamma }{\left( {\frac{{{n_b}}}{{{n_p}}}} \right)^{1/3}}{\left( {\cos {\theta _0}} \right)^{2/3}}{\left( {1 + {\gamma ^2}{{\tan }^2}{\theta _0}} \right)^{1/3}}.
\end{equation}
This implies that in the hydrodynamic regime, the fastest growing waves are oblique ones, i.e. waves at angle $\theta_0 \sim 1$. The growth rate for such waves is roughly 
 \begin{equation}
{\Gamma _{{\text{hydro}}}}\left(\theta_0 \sim 1\right) \sim {\omega _p}{\left( {\frac{{{n_b}}}{{\gamma {n_p}}}} \right)^{1/3}},
\end{equation}
while for waves which are almost parallel to the beam axis, i.e. $\theta_0 \sim \Delta\theta \sim \gamma^{-1}$, the growth rate reduces to 
\begin{equation}{\label{eq:GR_HYDRO2}}
{\Gamma _{{\rm{hydro}}}} \sim \frac{{{\omega _p}}}{\gamma }{\left( {\frac{{{n_b}}}{{{n_p}}}} \right)^{1/3}}.
\end{equation}
Therefore the beam evolution is dominated by oblique waves. In this case,
the beam angular spread will expand, without significantly losing energy \citep{Fainberg1970}. 

As the beam broadens, the transverse velocity dispersion, $\Delta v_{\perp} \sim c\Delta \theta$ increases accordingly and eventually the beam could not be considered `cold'. 
The hydrodynamic regime is valid as long as all the particles interact with the same wave, which implies:
\begin{equation}
    \Gamma \gg{{\bf{k}}\cdot\Delta {\bf{v}}},
\end{equation}
where $\Delta{\bf{v}}$ is the width of the particle velocity distribution. 
For a beam with angular spread $\Delta\theta$, the parallel and transverse velocity spreads are $\Delta v_{\perp} = c\Delta \theta$ and $\Delta v_{\parallel} = c{ \Delta\theta^2}/2$, respectively, so that the hydrodynamic regime condition reduces to
\begin{equation}
{\omega _p}\left( {\frac{{\Delta {\theta ^2}}}{2} + \tan {\theta _0}\Delta \theta } \right) \ll {\Gamma _{hydro}},
\end{equation}
where we have set the resonance condition, $k_{\parallel} = \omega_{p}/c$. 
Now one can write
\begin{equation}\label{eq:Hydro_Cond}
\begin{aligned}
& \Delta\theta \ll {\left( {\frac{{{n_b}}}{{\gamma {n_p}}}} \right)^{1/3}} & {;\quad {\theta _0} \sim 1} ,  \\ 
&   {\gamma\Delta\theta^2 \ll {\left( \frac{n_b}{n_p} \right)}^{1/3}} & {;\quad {\theta _0} \sim \Delta \theta }  .\\
\end{aligned}
\end{equation}
We show in section \ref{sec:IGM} that in agreement with \citet{Miniati2013}, the condition for the hydrodynamic regime is mostly violated for characteristic IGM and beam parameters so that the only applicable regime for the instability mechanism is the kinetic one.
\subsection{Kinetic regime}{\label{sec:KIN}}
The growth rate of a Langmuir wave, 
in the kinetic regime is given by (e.g., \citealt{BreizmanREVIEW})
\begin{equation}
\Gamma  = {\omega _p}\frac{{2{\pi ^2}{e^2}}}{{{k^2}}}\int {\left({\bf{k}} \cdot \frac{{\partial f}}{{\partial {\bf{p}}}} \right)\delta \left( {{\omega _p} - {\bf{k}} \cdot {\bf{v}}} \right){d^3}p} .
\end{equation}
Here the beam distribution function is normalized such that
\begin{equation}
\int {f{d^3}p}  = n_b.
\end{equation}
The beam is highly relativistic, which means that the magnitude of ${\bf{v}}$ is $c$ up to second order in $\gamma^{-1}$. Hence the delta function argument contains only the angles of the beam momentum vector. Therefore, integration of the delta function is performed over the azimuthal angle, $\phi$, and the growth rate function could be written as an integral solely over the polar angle. If one introduces the angular distribution function:
\begin{equation}{\label{eq:defG}}
g\left(\theta\right) =\int_0^{2\pi } {d\phi } \int_0^\infty  {f\left( {\mathbf{p}} \right)pdp}, 
\end{equation}
the growth rate can be written as 
\begin{equation}{\label{eq:GW2}}
\Gamma  = \frac{{{\omega _p}}}{2}\left( {\frac{{mc}}{{{n_b}}}} \right)\left( {\frac{{{n_b}}}{{{n_p}}}} \right){\left( {\frac{{{\omega _p}}}{{ck}}} \right)^3}\int_{{\theta _1}}^{{\theta _2}} {\frac{{\left[ {\frac{1}{\theta }\frac{{\partial g}}{{\partial \theta }}\left( {\cos \theta  - \frac{{ck}}{{{\omega _p}}}\cos {\theta _0}} \right) - 2g} \right]}}{{\sqrt {\left( {\cos {\theta _2} - \cos \theta } \right)\left( {\cos \theta  - \cos {\theta _1}} \right)} }}\theta d\theta } ,
\end{equation}
where the boundaries $\theta_{1,2}$ are given by:
\begin{equation}{\label{eq:GW_bound}}
\cos {\theta _{1,2}} = \frac{{{\omega _p}}}{{ck}}\cos {\theta _0} \pm \sin {\theta _0}\sqrt {1 - {{\left( {\frac{{{\omega _p}}}{{ck}}} \right)}^2}} .
\end{equation}
There is a simple estimate (\citealt{Breizman71}) for the maximal growth rate for a general wave vector angle, $\theta_0$:
\begin{equation}{\label{eq:GR_EST}}
\Gamma \sim \omega_{p} \left(\frac{n_b}{n_p}\right) \frac{\cos^2\theta_0}{\gamma\Delta\theta^2}.
\end{equation}
One sees that the growth rate 
decreases with increasing beam angular width, $\Delta\theta$.

The expression (\ref{eq:GR_EST}) is a good estimate but 
the exact dependence of the growth rate on $\theta_0$ is rather important since the growth of the oscillations is exponential: a factor of two in the growth rate value will eventually lead to difference by orders of magnitude between two waves' energies. The exact growth rate depends on the shape of the distribution function.

\begin{figure}
\centering
\includegraphics[width=\columnwidth]{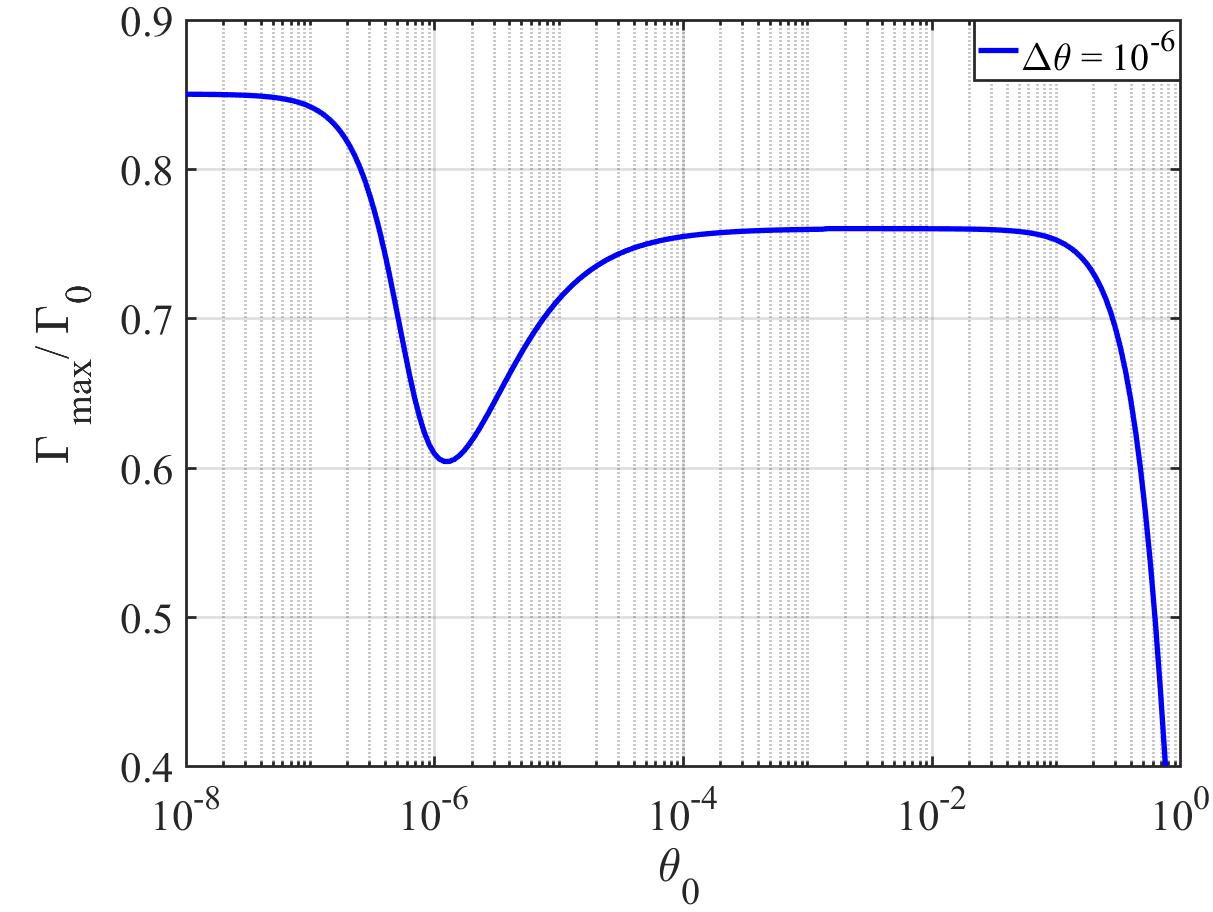}
\caption{The maximum growth rate, $\Gamma^{\text{max}}$, as a function of wave direction, $\theta_0$, for a Gaussian distribution with $\Delta\theta = 10^{-6}$. For any $\theta_0$, the maximum was found with respect to the resonance parameter $x$. The growth rate is normalized with respect to $\Gamma_0 = \omega_{p} \left(n_b/n_p\right) \left(\gamma \Delta\theta^2\right)^{-1}$.
}
\label{fig:MAXGrowthrate}
\end{figure}

To demonstrate the dependence of the growth rate on the wave vector, we consider a Gaussian distribution,
\begin{equation}
g(\theta ) = \frac{{{n_b}}}{{\gamma mc}}\frac{2}{{\Delta {\theta ^2}}}\exp \left( { - \frac{{{\theta ^2}}}{{\Delta {\theta ^2}}}} \right).
\label{Gauss}\end{equation}
In this case, the growth rate may be found in the form (see Appendix B)
\begin{equation}\label{eq:GR_GAUSS}
\begin{gathered}
  \Gamma  = {\omega _p}\left( {\frac{{{n_b}}}{{{n_p}}}} \right)\frac{{2\pi }}{{\gamma \Delta {\theta ^2}}}{\cos ^3}{\theta _0}\exp \left( { - \frac{{\theta _1^2 + \theta _2^2}}{{2\Delta {\theta ^2}}}} \right) \times  \hfill \\
  \left[ {\left( {{\xi ^2}{{\left( {\frac{{\Delta \theta }}{{2{\theta _0}}}} \right)}^2} - 1} \right){I_0}\left( \xi  \right) - \frac{\xi }{2}{I_1}\left( \xi  \right)} \right], \hfill \\ 
\end{gathered},
\end{equation}
where $\xi  \equiv \left( {\theta _2^2 - \theta _1^2} \right)/2\Delta {\theta ^2}$ and $I_{0,1}\left(x\right)$ are the modified Bessel functions of the first kind.
The growth rate depends on the direction of the wave, $\theta_0$, and on the absolute value of the wave vector, $k$. 
Figure \ref{fig:MAXGrowthrate} shows the maximum growth rate with respect to $k$ as a function of $\theta_0$. For large angles, $\theta_0\gg\Delta\theta$, the maximum growth rate varies as $\cos^2 \theta_0$, according to the general estimate (\ref{eq:GR_EST}). 
Because of that dependence, the growth rate is constant in a wide range of angles, $\Delta\theta\ll\theta_0\ll 1$.  
The shape of the maximum growth rate curve for $\theta_0\sim\Delta\theta$ is sensitive to the shape of the distribution function. 
Specifically for the Gaussian case, one gets a valley near $\theta_0 \sim \Delta \theta$ and the growth rate increases again to a plateau at $\theta_0 \ll \Delta \theta$. The plateau at very small angles is larger than the intermediate angles plateau $\Delta \theta \ll \theta_0 \ll 1$, for this case by roughly 10\%.

For every wave vector directed at the angle $\theta_0$, there exists a narrow range of $k$ which is within resonance with the narrow particle beam. One can conveniently use instead of $k$, the variable
\begin{equation}{\label{eq:DEFx}}
x = \frac{\frac{\omega_{p}}{ck} - \cos\theta_0}{\sqrt{1-\frac{\omega_p}{ck}\cos\theta_0}}.
\end{equation}
For the remainder of this analysis, we shall use the wave vector resonance parameter, $x$, as the wave parameter instead of $k$. The transition from $k$ to $x$ makes the various integration formulas more convenient. For large wave-vector angles, $\theta_0 \gg \theta$, 
\begin{equation}{\label{eq:RESX}}
\left|x\right| \le \theta,
\end{equation}
which is independent of $\theta_0$. For simple estimations, one can usually assume that $\left|x\right| \sim \Delta\theta$, where $\Delta\theta$ is the angular spread of the beam velocity distribution. Since $x$ is of the same order as $\theta$ and $\Delta\theta$, one can safely assume that $x$ is a small parameter, which 
considerably simplifies the derivation (see Appendix C). 

For $\Delta\theta \ll \theta_0$ and $\xi \gg 1$ 
equation (\ref{eq:GR_GAUSS}) is reduced to
\begin{equation}
\Gamma\left( {{\theta _0} \gg \left| x \right|} \right) =\omega_p \sqrt{\pi}\left( {\frac{{{n_b}}}{{{n_p}}}} \right)\left( {\frac{{ - x}}{{\Delta \theta }}} \right)\frac{{\exp \left( { - \frac{{{x^2}}}{{\Delta {\theta ^2}}}} \right)}}{{\gamma \Delta {\theta ^2}}}{\cos ^2}{\theta _0}.
\end{equation}
This growth rate is maximized for $x = -\frac{\Delta\theta}{\sqrt{2}}$ whose value is:
\begin{equation}{\label{eq:MAXGR}}
{\Gamma ^{{\rm{max}}}}\left( {{\theta _0} \gg \Delta\theta} \right) = {\omega _p}\sqrt {\frac{\pi }{{2e}}} \left( {\frac{{{n_b}}}{{{n_p}}}} \right)\frac{{{{\cos }^2}{\theta _0}}}{{\gamma \Delta {\theta ^2}}}.
\end{equation}
For the other extreme case where $\left| x\right | \gg \theta_0$ and $x<0$, which corresponds to very small wave vector angles, the growth rate is
\begin{equation}
\Gamma \left( {\left| x \right| \gg {\theta _0}} \right) = 2\pi {\omega _p}\left( {\frac{{{n_b}}}{{{n_p}}}} \right)\left( {\frac{{2{x^2}}}{{\Delta {\theta ^2}}} - 1} \right)\frac{{{e^{ -\frac{{2{x^2}}}{{\Delta {\theta ^2}}}}}}}{{\gamma \Delta {\theta ^2}}},
\end{equation}
which is maximized for $x = -\Delta \theta$ and its value is:
\begin{equation}
{\Gamma ^{{\rm{max}}}}\left( {{\theta _0} \to 0} \right) = \frac{{2\pi }}{{{e^2}}}{\omega _p}\left( {\frac{{{n_b}}}{{{n_p}}}} \right)\frac{1}{{\gamma \Delta {\theta ^2}}}.
\end{equation}
One can show that the modes $x>0$ are stable. We therefore do not consider positive values for $x$.

The behavior of the maximum curve at $\theta_0\sim\Delta\theta$ significantly depends on the shape of the distribution function (\citealt{Breizman71,Rudakov70}).
In the general case, the growth rate could only be found numerically. Taking into account a rather non-trivial behavior of the growth rate at $\theta_0\sim\Delta\theta$, we could not use the same general formula (\ref{eq:GW2}) and the same grid in order to find the growth rate in the entire range of possible resonance waves. Instead, we introduce three overlapping regions in the $(\theta_0,x)$ plane. Within each region we expand the general formula to leading order in the small parameters. After a long derivation (see Appendix D) we get:
\begin{enumerate}
\item {$\theta_0 \gg \left| x\right|$}
\begin{equation}
\Gamma  =
 \frac{{{\omega _p}}}{2}\left( {\frac{{mc}}{{{n_b}}}} \right)\left( {\frac{{{n_b}}}{{{n_p}}}} \right){\rm{sgn}}(x){\cos ^2}{\theta _0}\int_1^\infty  {\frac{{\frac{{\partial g}}{{\partial z}}}}{{\sqrt {z - 1} }}dz} ,
\end{equation}
where 
\begin{equation}
    z = \theta^2/x^2.
\end{equation}
Note that this function is anti-symmetric in $x$.
\item  {$\theta_0 \sim \left| x\right|$}
\newline
For brevity one can define new parameters: 
\begin{equation}
\mu \equiv \theta_0^2/x^2;\quad 
\rho \equiv {\mathop{\rm sgn}} (x)\sqrt {1 + 4\mu }  - 1.
\end{equation}
The boundaries of the integral, eq. (\ref{eq:GW_bound}), are rewritten in the variable $z$ as 
\begin{equation}
z_{1,2} = 2\mu  -\rho \pm 2  \sqrt {\mu\left(\mu  - \rho\right)} ,
\end{equation}
The resonance range of $x$ guarantees that the roots $z_{1,2}$ are real. The growth rate in this case is given by
\begin{equation}
\Gamma  = {\omega _p}\left( {\frac{{mc}}{{{n_b}}}} \right)\left( {\frac{{{n_b}}}{{{n_p}}}} \right)\int_{{z_1}}^{{z_2}} {\frac{{\frac{1}{2}\frac{{\partial g}}{{\partial z}}\left( {\rho  - z} \right) - g}}{{\sqrt {\left( {z - {z_1}} \right)\left( {{z_2} - z} \right)} }}dz} .
\end{equation}
\item {$\theta_0 \ll \left| x\right|$, $\quad x < 0$} 
\newline
For this limit, the integration boundaries approach the same value, so the integration effectively becomes a delta-integration at $z=2$. The growth rate reduces to a simpler form:
\begin{equation}
\Gamma  = \pi {\omega _p}\left( {\frac{{mc}}{{{n_b}}}} \right)\left( {\frac{{{n_b}}}{{{n_p}}}} \right){\left[ {\left( { - 2\frac{{\partial g}}{{\partial z}}} \right) - g} \right]_{z = 2}}.
\end{equation}
\end{enumerate}
The three cases show that the growth rate is always proportional to the same basic value $\Gamma_0$:
\begin{equation}
    \Gamma_0  =  {\frac{n_b}{n_p}} \frac{1}{\gamma\Delta\theta^2} {\omega _p},
\end{equation}
which is consistent with the estimate given by eq. (\ref{eq:GR_EST}).

\begin{figure}
    \centering
     \includegraphics[width=\columnwidth]{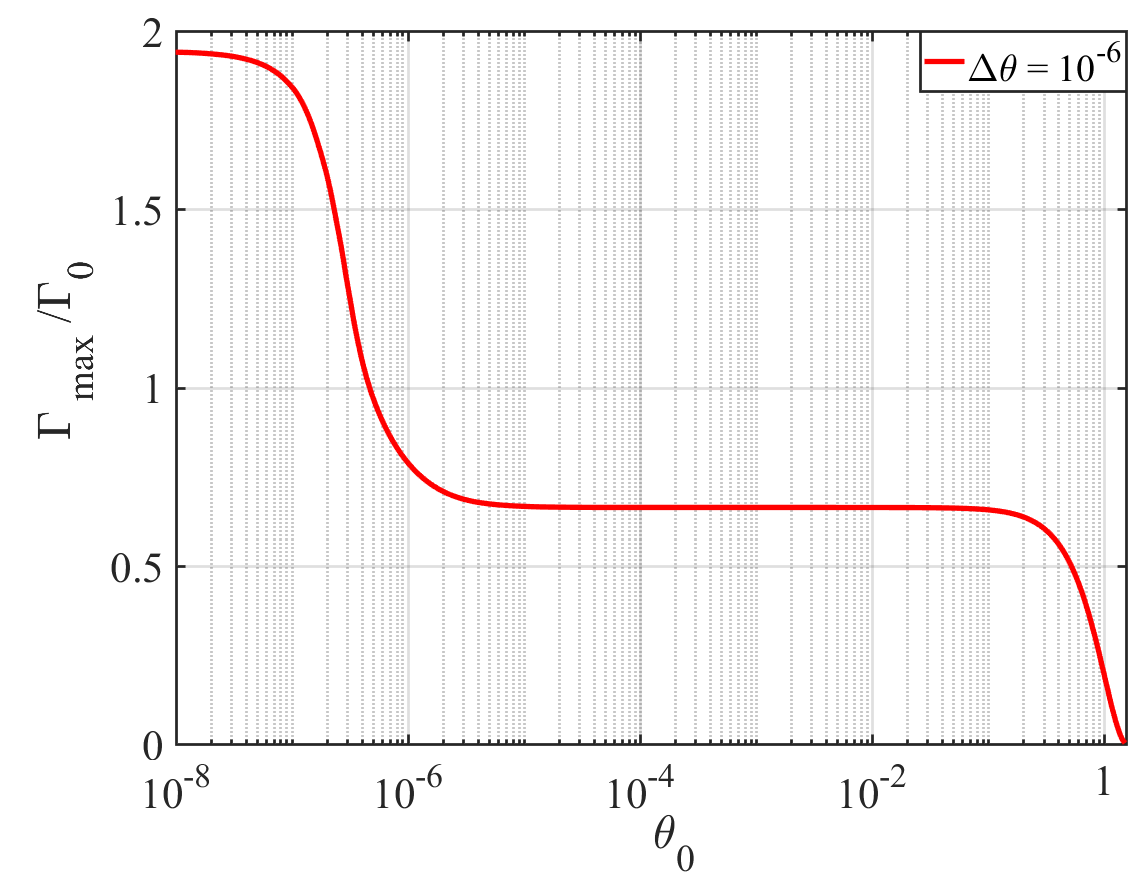}
    \caption {The same as in Fig.\ 2 
    for the beam distribution function (\ref{Rudakov_function}) 
    with $\Delta\theta = 10^{-6}$.}
    \label{fig:MAX_GR_DIST}
\end{figure}

As an example of a non-Gaussian angular distribution,  we calculate the instability growth rate for the function
\begin{equation} \label{Rudakov_function}
g = g_0 \exp{\left[ -\left(\frac{\theta}{\Delta \theta} \right)^5 \right ]},
\end{equation}
which describes the self-similar expansion of the beam \citep{Rudakov70}. 
Figure \ref{fig:MAX_GR_DIST} shows the maximal growth rate with respect to the variable $x$ as a function of $\theta_0$ for the case $\Delta\theta=10^{-6}$. 
One sees that for $\theta_0 \gg \Delta\theta$, the maximum growth rate is proportional to $\cos^2\theta_0$, according to the general estimate (\ref{eq:GR_EST}); 
this part is not sensitive to the shape of the distribution function. The effect of the beam distribution is seen clearly at
$\theta_0 \sim \Delta\theta$. In this range, the maximal growth rate increases by a factor of three. 
This is an effect of a steeper gradient in the distribution function at $\theta \sim \Delta\theta$. 
For distributions with a steep front, the growth rate for the angles $\theta_0 \sim\Delta\theta$ turns out to be larger than for moderate angles $\theta_0 \sim 1$ \citep{Rudakov70}. Hence the ratio between the two `plateau' values which are obtained for the maximal growth rate highly depends on the steepness of the distribution function. 

The larger growth rate at $\theta_0\sim\Delta\theta$ implies that the spectrum of excited waves is dominated by the waves propagating within the beam opening angle. These are the same waves which are capable of efficient energy exchange with the beam. This subtle fact explains why in the kinetic regime, relativistic beams efficiently lose their energy \citep{Fainberg1970,Breizman71,Rudakov70}. However, we have seen above that in very narrow beams, the width of resonance for these waves is especially low. Therefore in this crucial range of angles, the instability is easily suppressed. Below we show that no such condition exists which allows the excitation of these waves by relativistic intergalactic beams.

\section{Estimates for intergalactic beams}{\label{sec:IGM}}
Now let us apply these general results to extragalactic beams in particular. 
The IGM  plasma density is \citep{Broderick12}
\begin{equation}{\label{eq:np}}
    n_{p} = 2.2 \times 10^{-7} 
    \left(1+z\right)^3 \text{cm}^{-3},
\end{equation}
where $z$ is the redshift of the TeV blazar. 
The beam density $n_b$ is determined by 
balance between cooling and creation processes.
The upper limit for the beam's density is found when the cooling is determined by inverse Compton (IC) scattering. The IC scattering rate is given by \citep{Broderick12}:
\begin{equation}
    {\Gamma _{{\text{IC}}}} = 1.4 \times {10^{ - 20}}{\left( {1 + z} \right)^4}\gamma \quad {{\text{s}}^{ - 1}},
\end{equation}
and the corresponding upper limit for the beam density derived from the cooling rate is:
\begin{equation}{\label{eq:MAX_NB}}
 n_{b}^{\text{max}} \sim 5.1 \times 10^{-25} \left({1+z}\right)^{9.5} \left(\frac{L_{E}}{10^{45} \text{erg} \; \text{s}^{-1}}\right) \left(\frac{E}{\rm TeV}\right) \text{cm}^{-3}.
\end{equation}
Here $L_{E}$ is the gamma-ray luminosity of the blazar at the photon energy $E$. The Lorenz factors of the produced pairs are found in the range $\gamma\sim 10^4-10^7$  with an average of $\gamma\sim 10^5$ \citep{Miniati2013}. The pairs are produced at an angular spread $\Delta\theta\sim \gamma^{-1}$.

Substituting the above plasma and beam densities into the condition for the hydrodynamic instability regime, Eq.\  (\ref{eq:Hydro_Cond}), and taking into account that 
$\Delta\theta\sim \gamma^{-1}$, one finds that the hydrodynamic regime is valid only for large Lorentz factor values:
\begin{equation}
\begin{aligned}
& \gamma  \gg 7.5 \times {10^7}{\left(1 + z \right)^{-13/6}}
{\left( {\frac{L_{E}}{{{{10}^{45}}{\text{erg}}\;{{\text{s}}^{ - 1}}}}} \right)^{ - 1/3}}  & {;\quad {\theta _0} \sim 1} ,  \\ 
&\gamma  \gg 8 \times {10^5}{\left(1 + z \right)^{-13/8}}
{\left( {\frac{L_{E}}{{{{10}^{45}}{\text{erg}}\;{{\text{s}}^{ - 1}}}}} \right)^{ - 1/4}}  & {;\quad {\theta _0} \sim \Delta \theta }  .\\
\end{aligned}
\end{equation}
One sees that the condition for the hydrodynamic regime is not satisfied for oblique waves and could only be marginally satisfied for quasi-parallel waves. This means that
the treatment of the beam instability in the kinetic regime is justified for typical beams in the IGM.
Now let us estimate the role of plasma inhomogeneity on the development of the instability. It was discussed in section\ \ref{sec:inhomogeneity} that the effects of plasma inhomogeneities on the instability result in loss of resonance between the particles and the Langmuir wave caused by changes in the wave vector at a rate given by (\ref{k-drift}). 
The resonant wave grows exponentially with a rate $\Gamma$ from the initial thermal noise level. In order for the wave to grow significantly, the exponent has to be rather large. Therefore the time of growth must exceed $t=\Lambda/\Gamma$, where $\Lambda$ is the Coulomb logarithm. 

It was shown in section \ref{sec:resonance} that the resonance width of wave numbers strongly depends on the angle between the wave and the beam axis, $\theta_0$, and may be estimated as
\begin{equation}
    \frac{\delta k}k\sim 
    \left\{\begin{array}{ll}  \Delta\theta; & \theta_0\sim 1\\ (\Delta\theta)^2; & \theta_0\sim\Delta\theta.
\end{array}\right.
\end{equation}
Now the condition for the development of the instability can be written as:
\begin{equation}
\frac{\Lambda c}{L\Gamma}\leq\frac{\delta k}k\sim \left\{\begin{array}{ll}  \Delta\theta; & \theta_0\sim 1\\ (\Delta\theta)^2; & \theta_0\sim\Delta\theta.
\end{array}\right.
\end{equation}
Making use of the estimate (\ref{eq:MAXGR}) for the kinetic growth rate 
and introducing the following short-hand notations:
\begin{equation}
\nonumber
 n_{b} = n_{b,-22} \times 10^{-22} \text{cm}^{-3} \: , \:
  n_{p} = n_{p,-7} \times 10^{-7} \text{cm}^{-3},
\end{equation}
we can write the condition in the form:
\begin{eqnarray}{\label{eq:INHOM_cond}}
  0.1\left( {\gamma \Delta \theta } \right)\left( {\frac{\Lambda }{{20}}} \right)\left( {\frac{{n_{p, - 7}^{1/2}}}{{{n_{b, - 22}}}}} \right)\left( {\frac{{100Mpc}}{L}} \right)<  1;\qquad \theta_0\sim 1\\  0.1\gamma\left( {\frac{\Lambda }{{20}}} \right) \left( {\frac{{n_{p, - 7}^{1/2}}}{{{n_{b, - 22}}}}} \right)\left( {\frac{{100Mpc}}{L}} \right)<1;\qquad \theta_0\sim\Delta\theta.
\end{eqnarray}
We normalize $L$ by the maximal inhomogeneity scale, of the order of the cosmological void side. One can see that even in this case, only oblique waves could grow. Interaction of the beam with these waves leads only to expansion of the beam without noticeable loss of energy. The instability ceases when the beam expands about an order of magnitude, $\gamma\Delta\theta\sim 10$. Nearly parallel modes, $\theta_0\sim\Delta\theta$, which could take the energy off the beam, 
would not grow at the relevant Lorentz factors of the beam, $\gamma\sim 10^4-10^7$. 
These are the only waves which
could take a significant fraction of the beam energy, but one can see that their growth is completely suppressed by the inhomogeneity condition. Therefore even the weakest possible inhomogeneity makes any energy loss of the beam inefficient.

Now let us come back to the theory proposed by \citet{shalaby2018growth,Shalaby2020}, which is discussed in sect.\ 4. Even if their theory were universlly correct, their instability growth rate (see  eq.\ 4.16 in \citealt{Shalaby2020}) would only be
\eqb
\Gamma\sim\frac 12\left(\frac{2n_b}{n_p\gamma^3}\right)^{2/5}\omega_p= 7.5\times 10^{-13}\frac{n^{0.4}_{b,-22}n^{0.1}_{p,-7}}{\gamma^{1.2}_6}\,\rm s^{-1},
\eqe
which is less rapid than the collisional decay time of the plasma waves (e.g., \citealt{Miniati2013}) 
\begin{equation}
{\label{eq:Coulomb}}
    \nu_{\rm coll}\approx 10^{-11}n_{p,-7}T_4^{-3/2}
    \,\rm s^{-1},
\end{equation}
where $T=T_4\times 10^4$ K.
We stress out that their result was obtained in the hydrodynamic approximation  of the instability, even though it should be described by kinetic theory at these parameters. Then the kinetic instability is weaker than the hydrodynamic one at the same beam density because in the kinetic regime, only a fraction of electrons resonantly excite the wave. This means that the waves propagating along the beam are not excited in any case. As to oblique waves, it was explained above and will be confirmed by the rigorous theory in the next section that the excitation of these waves results merely in the expansion of the beam, but doesn't lead to energy loss. In the next section, we demonstrate straightforwardly that even if the oblique instability develops, the beam just expands, while energy loss remains negligibly small.

\section{Evolution of the beam; quantitative theory}
\subsection{Particle diffusion in momentum space}
When excited waves' energy is small in comparsion with the plasma's thermal energy, the evolution of the beam distribution function is governed by the diffusion equation in momentum space (e.g. \citealt{Breizman71, BreizmanREVIEW})
\begin{equation}
\frac{\partial f}{\partial t} 
= \frac{\partial }{\partial p_{\alpha}} \left( D_{\alpha \beta} \frac{\partial f}{\partial p_{\beta}} \right),
\end{equation}
where $D_{\alpha \beta}$ is the resonant momentum-diffusion tensor defined by
\begin{equation}
{D_{\alpha \beta }} = {\pi e^2}\int {W\left( {{\bf{k}},t} \right)\frac{{{k_\alpha }{k_\beta }}}{{{k^2}}}\delta \left( {{\bf{k}} \cdot {\bf{v}} - \omega } \right){d^3}k} .
\end{equation}
For the spherical, azimuthal-symmetric case, 
the diffusion equation is written as (under the previous assumption that the beam is narrow, $\theta\ll1$)
\begin{equation}{\label{eq:Diff_eq}}
\begin{array}{l}
\frac{{\partial f}}{{\partial t}}
= \frac{1}{{{p^2}\theta }}\frac{\partial }{{\partial \theta }}\left( {\theta {D_{\theta \theta }}\frac{{\partial f}}{{\partial \theta }}} \right) + \frac{1}{{p\theta }}\frac{\partial }{{\partial \theta }}\left( {\theta {D_{\theta p}}\frac{{\partial f}}{{\partial p}}} \right)\\
 + \frac{1}{{{p^2}}}\frac{\partial }{{\partial p}}\left( {p{D_{\theta p}}\frac{{\partial f}}{{\partial \theta }}} \right) + \frac{1}{{{p^2}}}\frac{\partial }{{\partial p}}\left( {{p^2}{D_{pp}}\frac{{\partial f}}{{\partial p}}} \right).
\end{array}
\end{equation}
The diffusion coefficients become
\begin{equation}\label{eq:DIFF_COEFF}
{D_{\mu }} = {\pi e^2}\int {W\left( {{\bf{k}},t} \right)\left(\frac{k_{\theta}}{k}\right)^{\mu}\delta \left( {{\bf{k}} \cdot {\bf{v}} - \omega } \right){d^3}k} ,
\end{equation}
where $\mu = 0,1,2$ for $D_{pp},D_{\theta p},D_{\theta \theta}$, respectively.

It follows immediately from eq. (\ref{eq:DIFF_COEFF}), that if the spectral energy of oscillations, $W\left({\bf{k}}\right)$, is dominated by waves propagating within the angle $\theta_0$, 
the diffusion coefficients are related as
\begin{equation}
    {D_{\theta \theta }}:{D_{\theta p }}:{D_{pp}}\sim \theta_0^2:\theta_0:1.
\end{equation}
The intergalactic beams are extremely narrow, $\Delta\theta\sim 10^{-5}$, whereas their energy spread is large $\Delta p\sim p$. In this case, the different terms in the R.H.S. of eq.\ (\ref{eq:Diff_eq}) are related as
\begin{equation}\label{eq:diffusion_ratio}
\begin{gathered}
  \frac{1}{\theta }\frac{\partial }{{\partial \theta }}\left( {\theta {D_{\theta \theta }}\frac{{\partial f}}{{\partial \theta }}} \right):\frac{p}{\theta }\frac{\partial }{{\partial \theta }}\left( {\theta {D_{\theta p}}\frac{{\partial f}}{{\partial p}}} \right) \hfill \\
  :\frac{\partial }{{\partial p}}\left( {p{D_{\theta p}}\frac{{\partial f}}{{\partial \theta }}} \right):\frac{\partial }{{\partial p}}\left( {{p^2}{D_{pp}}\frac{{\partial f}}{{\partial p}}} \right) \hfill \\
 \sim{\left( {\frac{{{\theta _0}}}{{\Delta \theta }}} \right)^2}:\left( {\frac{{{\theta _0}}}{{\Delta \theta }}} \right):\left( {\frac{{{\theta _0}}}{{\Delta \theta }}} \right):1 \hfill \\ 
\end{gathered} 
\end{equation}
One sees that if the energy of the excited waves is dominated by oblique waves, $\theta_0\gg\Delta\theta$, the first term in the R.H.S. of eq. (\ref{eq:Diff_eq}) significantly exceeds the others, which means that the beam expands without losing energy. Only if the waves are excited predominantly within the beam opening angle, $\theta_0\sim\Delta\theta$, all the terms on the R.H.S. are comparable, and then the beam's energy is efficiently redistributed.  

A special feature of the kinetic beam instability is that at $\theta_0<\Delta\theta$, the growth rate is larger than at a larger $\theta_0 \sim \Delta \theta$ (\citealt{Breizman71,Rudakov70}, see also section\ 3.2).  Due to the exponential growth from a very low level of initial thermal fluctuations, nearly parallel waves could reach a much larger energy than oblique ones. This subtle fact leads to the widely known statement \citep{Fainberg1970, Breizman71, Rudakov70} that in the kinetic regime, relativistic beams efficiently lose their energy. However, the width of resonance for these waves is extremely small. Therefore the instability of these waves may be easily suppressed by, e.g., weak inhomogeneity (\citealt{Breizman70}, see also section\ 2.2). The estimates show (see section\ 4) that for typical parameters of intergalactic beams, this effect completely suppresses the excitation of nearly parallel waves, so the beam could only expand without cooling. In section \ref{sec:SIMUL} we present numerical simulations supporting the estimates above.

\subsection{Evolution of the wave spectrum}
The evolution equation for the oscillation spectrum density, $W\left(\bf{k}\right)$, is written in the quasilinear approximation as
\begin{equation}
    \frac{\partial W}{{\partial t}} + {\bf v}_g \cdot \nabla W - \nabla \omega  \cdot \frac{{\partial W}}{{\partial {\bf{k}}}} = 2\left(\Gamma-\nu_{\rm coll}\right)W,
\label{eq:dW/dt}\end{equation}
where  $\nu_{\rm coll}$ 
is the collisional decay rate of the plasma waves (\ref{eq:Coulomb}).
The l.h.s. of this equation is the full time derivative of the plasma wave spectral energy density along the rays which obey the geometrical optics law:
\begin{equation}
    \frac{d\bf r}{dt}=\frac{{\partial \omega }}{{\partial {\bf{k}}}};\qquad
    \frac{d\bf k}{dt}=-\nabla\omega.
\label{optics}\end{equation}
The first term in (\ref{eq:dW/dt}) represents the time evolution, the second is spatial advection, and the third term describes a spectral shift in $k$-space due to inhomogeneity. 
The r.h.s. of this equation describes the growth or damping 
of waves' energy due to resonance interaction with the beam and decay due to Coulomb collisions (see Eq.\ (\ref{eq:Coulomb})). 

We do not take into account non-linear processes such as induced scattering and modulation instability of the plasma waves.
These processes redistribute the energy of the waves over the spectrum, thus removing them from the resonance region. Therefore they provide additional mechanisms for the stabilization of the beam instability. We show that the inhomogeneity solely restricts the efficiency of energy loss via the beam instability. Thus, there is no special need to consider additional processes.

 A formal solution to  eq.\ (\ref{eq:dW/dt}) is
\begin{equation}
    W(t,{\bf r},{\bf k})=W_0\exp\left\{
    2\int_0^t\left[\Gamma(t',{\bf r}',{\bf k}')-\nu_{\rm coll}({\bf r}')\right]dt'\right\},
\label{eq:W_solution}\end{equation}
where the integral is along the characteristics ${\bf r}'={\bf r}+{\bf v}_g(t'-t)$, ${\bf k}'={\bf k}-\nabla\omega (t'-t)$. The growth rate $\Gamma$ is given by eq. (\ref{eq:GW2}). The time-dependence of the growth rate, $\Gamma$, is determined by the time variation of the distribution function. The dependence on $\bf k$ is very strong because the resonance range $\Delta k$ is very narrow.
Both $\Gamma$ and $\nu_{\rm coll}$ depend on the spatial coordinate only via the plasma density, which enters the expressions for $\Gamma$ and $\nu_{\rm coll}$ as  a multiplier and varies at a very large scale $L$. Taking into account that the light travel time across the spatial inhomogeneity scale significantly exceeds the instability time, $t\ll L/c$, one can neglect this dependence in the integrand. Assuming that the initial fluctuation rate, $W_0$, is spatially homogeneous, one finds that the dependence on $\bf r$, and therefore on ${\bf v}_g$, drops out of the solution. This means that one can drop the second term in the l.h.s. of eq.\ (\ref{eq:dW/dt}): the evolution of the wave density is determined by the time evolution of $\bf k$ according to the second equation in (\ref{optics}) and not by the spatial transfer.

For numerical purposes, we will use  eq.\ (\ref{eq:dW/dt}) but not eq. (\ref{eq:W_solution}).
\newline
Now we can rewrite this equation in a simplified form:
\begin{equation}
\frac{{\partial W}}{{\partial t}} + \frac{\omega_p}{L} \frac{{\partial W}}{{\partial {k }}} = 2\left(\Gamma -\nu_{\rm coll}  \right)W.
\end{equation}
Taking into account that the waves are concentrated in a very narrow range of $k$, one can conveniently rewrite, for numerical purposes, the equation in terms of the resonant parameter (\ref{eq:RESX}), which is roughly bounded by $\vert x\vert\le\Delta\theta$. Then
for the case $\theta_0 \gg\Delta\theta$, we get
\begin{equation}{\label{eq:WOSC_EQ1}}
\frac{{\partial W}}{{\partial t}}  - \frac{c}{L}\frac{{{{\cos }^2}{\theta _0}}}{{\sin {\theta _0}}}\frac{{\partial W}}{{\partial x}} = 2\left(\Gamma -\nu_{\rm coll}  \right) W;
\end{equation}
whereas for 
$\theta_0\sim\Delta\theta$, the equation may be written as
\begin{equation}
\frac{{\partial W}}{{\partial t}} - \frac{2}{{{\theta ^2}}}\frac{c}{L}\frac{{\partial W}}{{\partial y}} = 2\left(\Gamma -\nu_{\rm coll}  \right)W,
\label{eq:Wosc2}\end{equation}
where 
\begin{equation}
y = \frac{{x\sqrt {{x^2} + 4\theta _0^2}  - {x^2}}}{{{\theta ^2}}} + 1 .
\end{equation}
One can show 
that the resonance width $\Delta y$ is roughly $\theta_0/\theta$ so that $\Delta y\sim 1$ for $x \sim \theta$.

\subsection{Numerical simulation}{\label{sec:SIMUL}}
In this section we present the results of our numerical simulation of the beam-plasma interaction in the quasi-linear regime.

The extragalactic beams are extremely narrow, $\Delta\theta\sim 10^{-5}$. Our previous estimates show that only oblique waves, $\theta_0\sim 1$, are excited, in the presence of inhomogeneity. This implies that the first term on the R.H.S. of the momentum diffusion equation (\ref{eq:Diff_eq}) exceeds other terms by 5-10 orders of magnitude. Therefore we initially only consider the evolution of the angular distribution of particles according to the equation
\begin{equation}\label{eq:DIFFUSION}
\frac{{\partial g}}{{\partial t}} = \frac{1}{{{{\left( {\gamma mc} \right)}^2}}}\frac{\partial }{{\partial {\theta ^2}}}\left( {4{\theta ^2}{D_{\theta \theta }}\frac{{\partial g}}{{\partial {\theta ^2}}}} \right).
\end{equation}
where the angular distribution function, $g$, has been defined in ($\ref{eq:defG}$).
Having found the angular evolution of the beam, together with the spectral evolution of the wave, we evaluate the other diffusion coefficients in the momentum diffusion equation. If they are comparable to $D_{\theta\theta}$, then according to our estimates the other diffusion processes are negligible, which is self-consistent with our assumption.

For each time step, we first solve the diffusion equation (\ref{eq:DIFFUSION}) given the diffusion coefficients from the last time step. After the distribution function is updated, we calculate the growth rates for each wave vector, making use of the formulas presented in section \ref{sec:KIN}.
After the growth rates are calculated in the current time step, we next calculate the new values of the oscillations energy density, $W$, according to the evolution equation (eq. \ref{eq:WOSC_EQ1} or, for small $\theta_0$, \ref{eq:Wosc2}).
Using the updated values of $W$, we calculate the diffusion coefficients (the procedure is described in Appendix E).
The new values of the diffusion coefficients are used in the next time step for a new cycle as described above. 

Each separate simulation is characterized by a set of $\left(\gamma,n_{p},n_{b},L\right)$.
The simulation time is measured in units of the characteristic instability time 
\begin{equation}
    t_{0} = \Gamma_{0}^{-1} = \left(n_p/n_b\right)\omega_{p}^{-1} \gamma^{-1}.
\label{eq:inst_time}\end{equation}
The initial angular distribution function is chosen in the form 
\begin{equation}{\label{eq:INIT_DIST}}
    g\left(\theta\right) = \exp\left(-0.2 \left(\theta\gamma\right)^5\right) ,
\end{equation}
so that the initial angular spread is always $\Delta \theta \approx \gamma^{-1}$, and the function has a sharp gradient around $\theta \sim \gamma^{-1}$.
The oscillations energy density is initialized uniformly to a low level such that the angular diffusion rate due to the initial background is negligibly slow (see below). 
In all of the simulations we performed, we have fixed the value of the plasma density to be $n_p = 2.2 \times 10^{-7} \text{cm}^{-3}$, which corresponds to no redshift. 
We consider 
the maximal possible plasma inhomogeneity length scale, $L = 100Mpc$, providing the minimal influence of inhomogeneity. 

\begin{figure}
    \centering
    \includegraphics[width=\columnwidth]{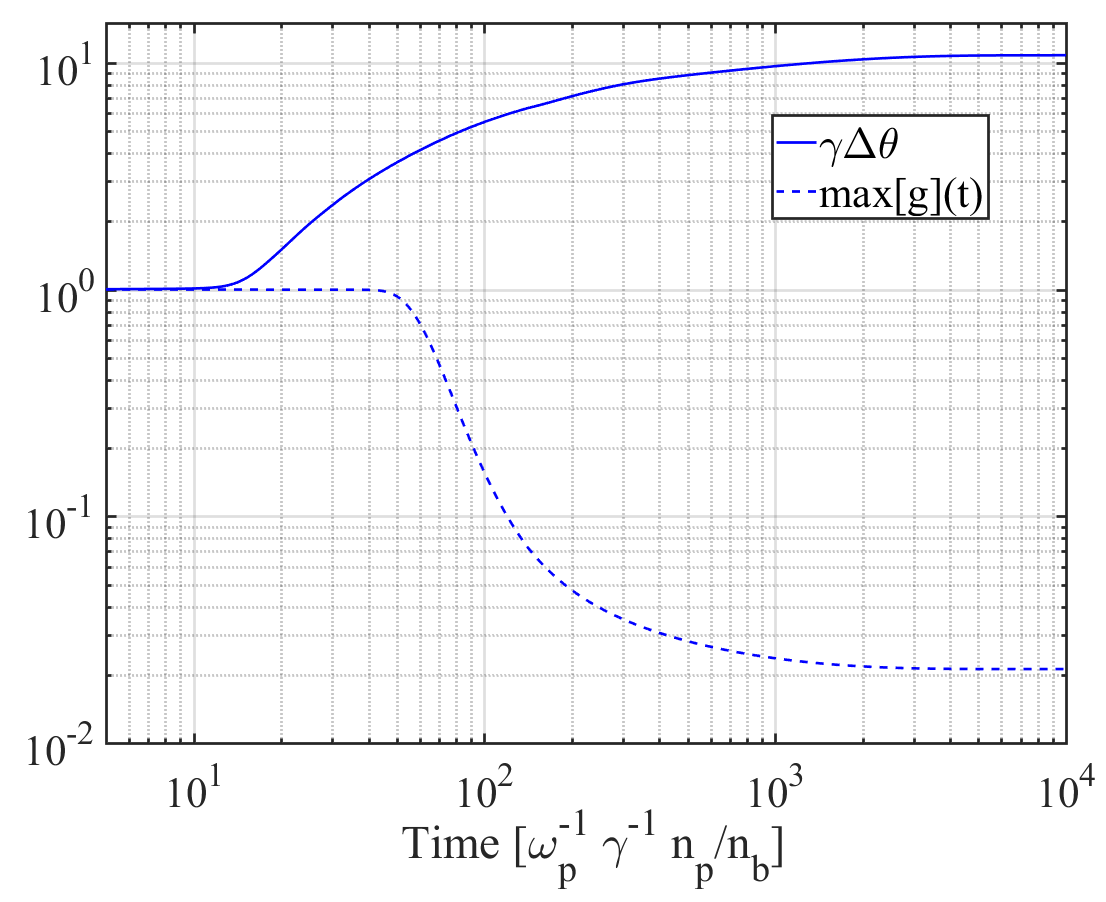}
    \caption{The angular spread of the beam distribution function (solid) as a function of normalized time and $\max \left(g\right)$ (dashed) as a function of normalized time. $\gamma = 10^{6} $, $n_{b,-22} = 1$, $L = 100Mpc$.}
   \label{fig:MAXG_DTHETA}
\end{figure}

Figure \ref{fig:MAXG_DTHETA} shows the angular spread, $\Delta\theta$, and the maximum value of the angular distribution function, ${\rm max} \left[g\right]$, as a function of the normalized time. The beam is spreading in angle while the distribution flattens, which means that particles gain transverse momentum and they are drifting away from the initial direction of the beam. The spreading begins after about a dozen instability times, when the resonant wave grew enough from the initial low level. After the beam spreads by about an order of magnitude, the growth rate of the  instability falls below the collision decay rate of the plasma waves, and the process is saturated.

\begin{figure}
    \centering
    \includegraphics[width=\columnwidth]{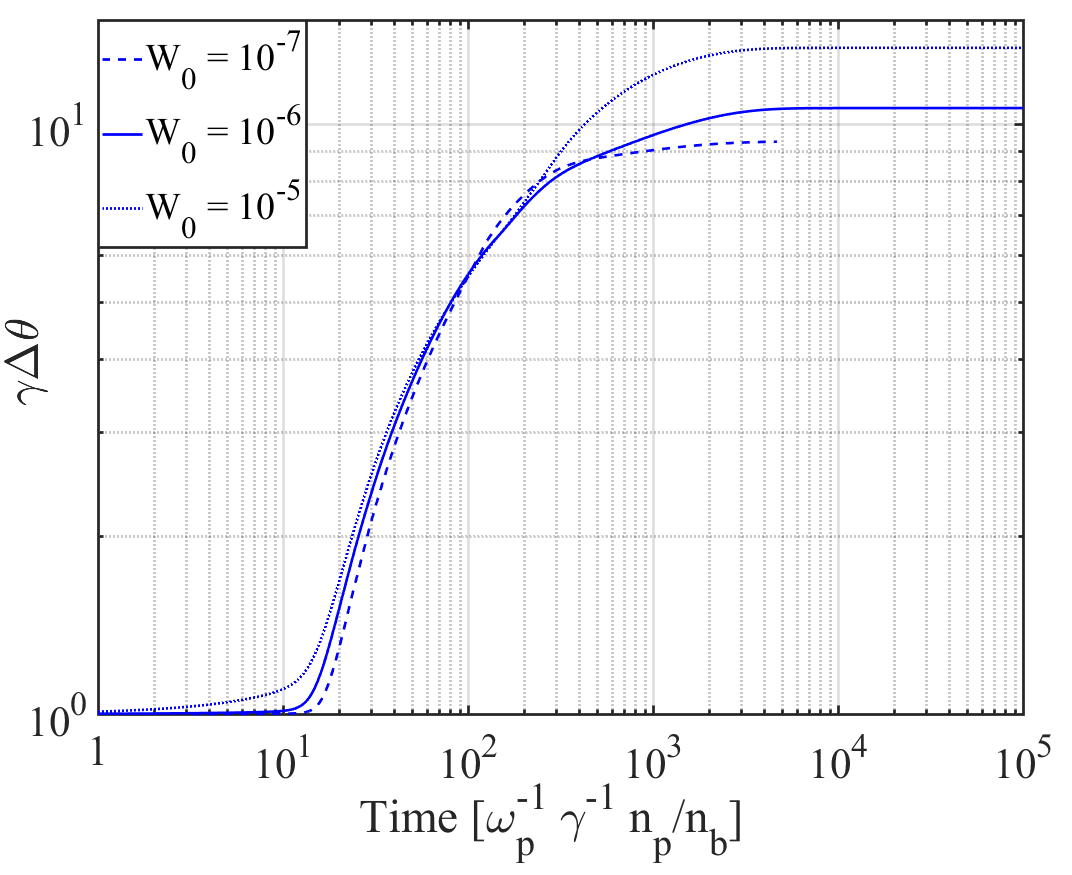}
    \caption{The normalized angular spread, $\gamma\Delta\theta$, as a function of normalized time for different initial levels of oscillations, $W_0$, see eq.\ (\ref{W_unit}). The beam parameters are the same as in the previous figure. 
    }
   \label{fig:DTHETA_NBS}
\end{figure}

In figure \ref{fig:DTHETA_NBS} we present simulations of the beam with the same parameters but with different initial plasma wave energy levels, $\tilde{W}_0$, normalized by \begin{equation}{\label{W_unit}}
    \tilde{W}_0 =\frac{mc^2}{\omega_p t_0} \left(\frac{e^2}{mc^2} \frac{\omega_p}{c}\right)^{-1} W_0,
\end{equation}
where $W_0$ is dimensionless. One sees that the results are weakly dependent on $W_0$, the initial level of oscillations, provided it is small enough.

\begin{figure}
    \centering
    \includegraphics[width=\columnwidth]{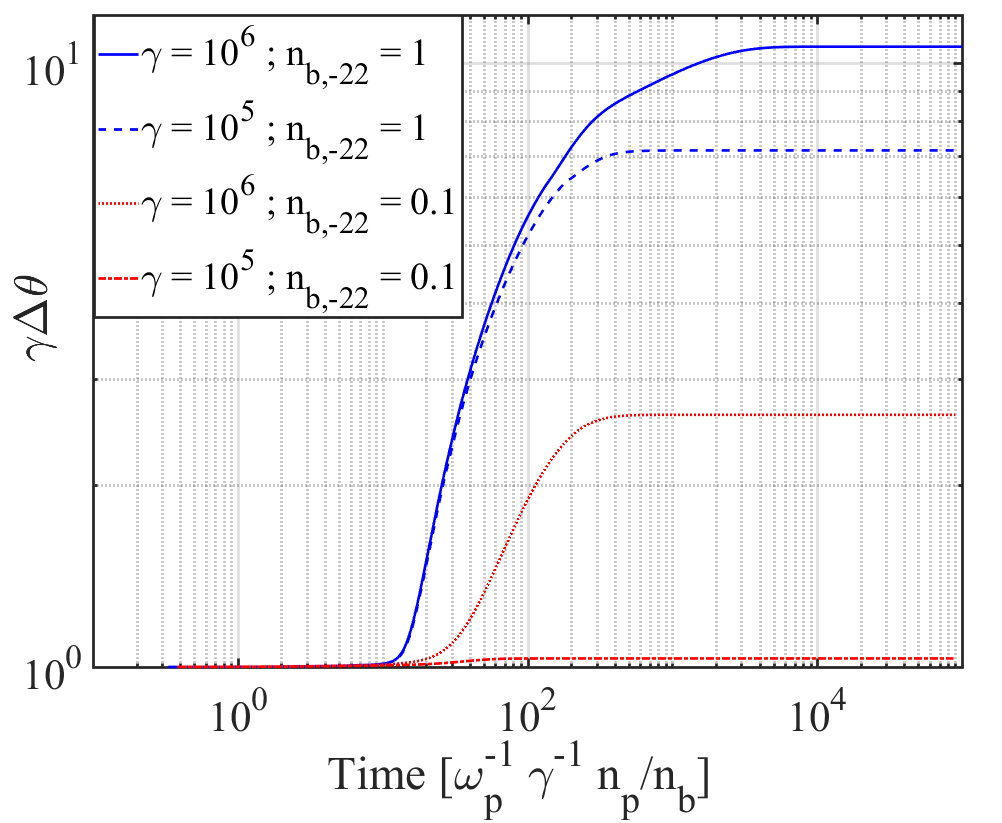}
    \caption{The normalized angular spread, $\gamma\Delta\theta$ as a function of normalized time for $\gamma = 10^{6}, n_{b,-22} = 1$ (solid line), $\gamma = 10^{5}, n_{b,-22} = 1$ (dashed line), $\gamma = 10^{6}, n_{b,-22} = 0.1$ (dotted line), $\gamma = 10^{5}, n_{b,-22} = 0.1$ (dashed-dotted line). $L = 100Mpc$.}
   \label{fig:DTHETA_GAMMAS}
\end{figure}
Figure \ref{fig:DTHETA_GAMMAS} shows the evolution of the beam width for different values of beam density and Lorentz factors. One sees that in all cases, the behavior of the beam is essentially the same: the expansion saturates after the beam expands a few times over. This agrees with the result of our qualitative estimates presented in section 4. 

\begin{figure}
    \centering
    \includegraphics[width=\columnwidth]{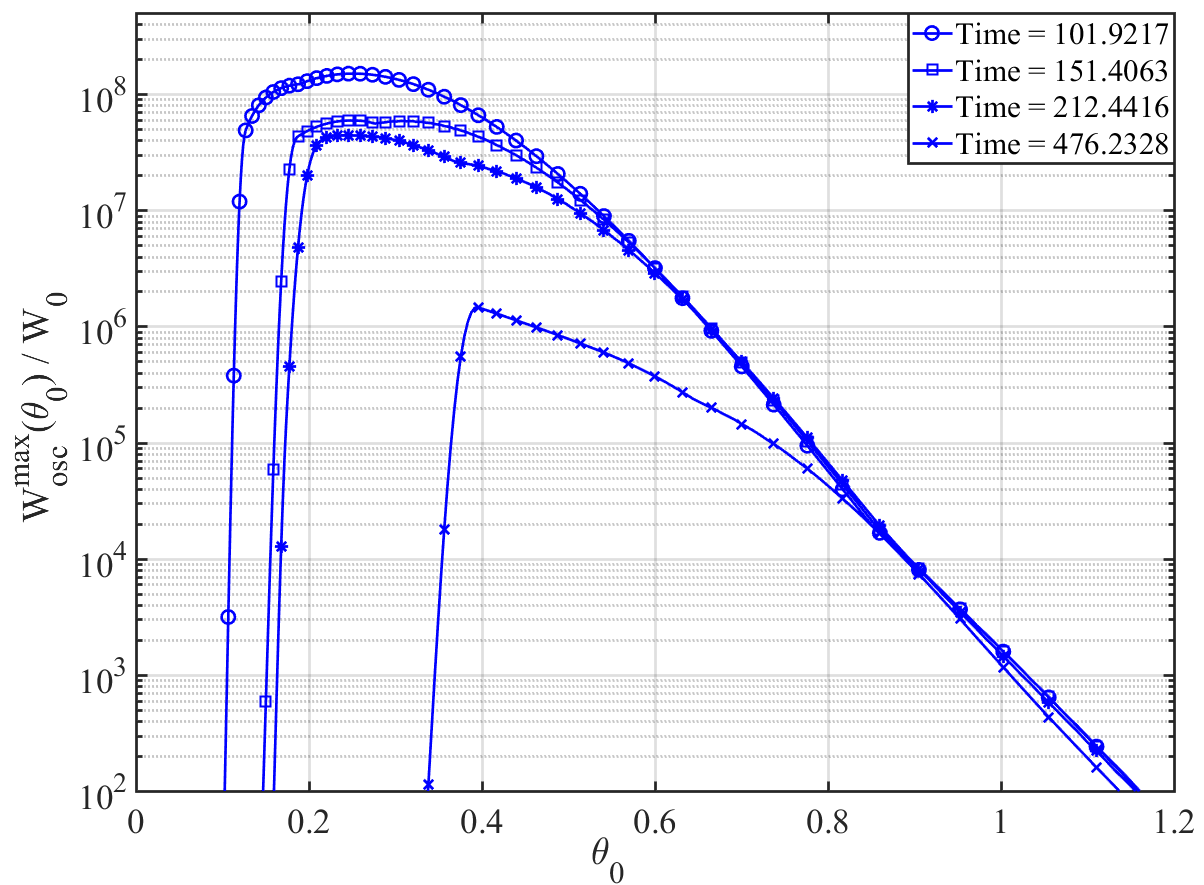}
    \caption{The maximum oscillation energy density $W_{\rm osc}$ with respect to the resonance parameter $x$ as a function of $\theta_0$ for different times (normalized by the instability time (\ref{eq:inst_time})). $n_{b,-22} = 1$ ,$ \gamma= 10^{6}$, $L = 100Mpc$.}
    \label{fig:WOSC_nb24}
\end{figure}

\begin{figure}
    \centering
    \includegraphics[width=\columnwidth]{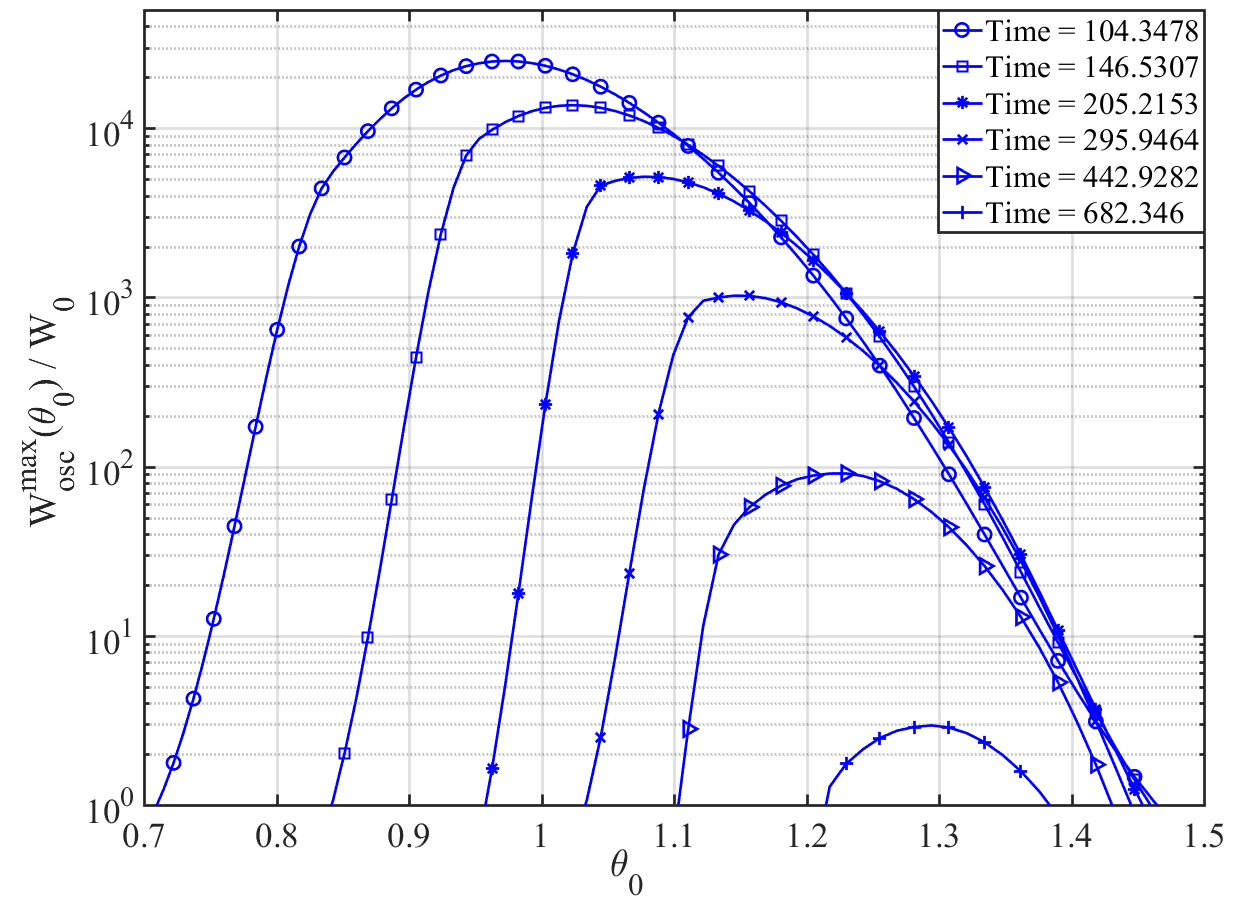}
    \caption{Same as figure \ref{fig:WOSC_nb24}, but with $n_{b,-22} = 10^{-1}$.}
   \label{fig:WOSC_nb23}
\end{figure}

In our numerical simulation, we calculate the growth and decay of the resonant Langmuir waves energy.
Figures \ref{fig:WOSC_nb24} and \ref{fig:WOSC_nb23} show the maximum oscillation energy density with respect to $k$ as a function of $\theta_0$ at different times. We measure the oscillations energy density with respect to their initial value, $W_0$ which is considered as the dimensionless noise level. The first figure shows results for $n_{b} = 10^{-22} \text{cm}^{-3}$ and the second for  $n_{b} = 10^{-23} \text{cm}^{-3}$. 
One sees that each curve has a maximum at some angle $\theta_0 \sim 1$. This peak is a result of the competition between two terms: the growth, which is proportional to $\cos^2\theta_0$, thus stronger for smaller angles, and the inhomogeneity term, which suppresses the instability more efficiently for smaller angles.
Both cases show that small angles waves do not grow, as expected. 
Note that whereas the densities of two beams differ by a factor of 10, 
the level of the oscillations energy differs by four orders of magnitude. The reason is that the instability growth rate for the smaller density beam is ten times smaller than that of the denser beam. Therefore for the more dilute beam, inhomogeneity also succeeds in suppressing the growth of oblique waves, $\theta_0\sim 1$ (see below).

\begin{figure}
    \centering
    \includegraphics[width=90mm]{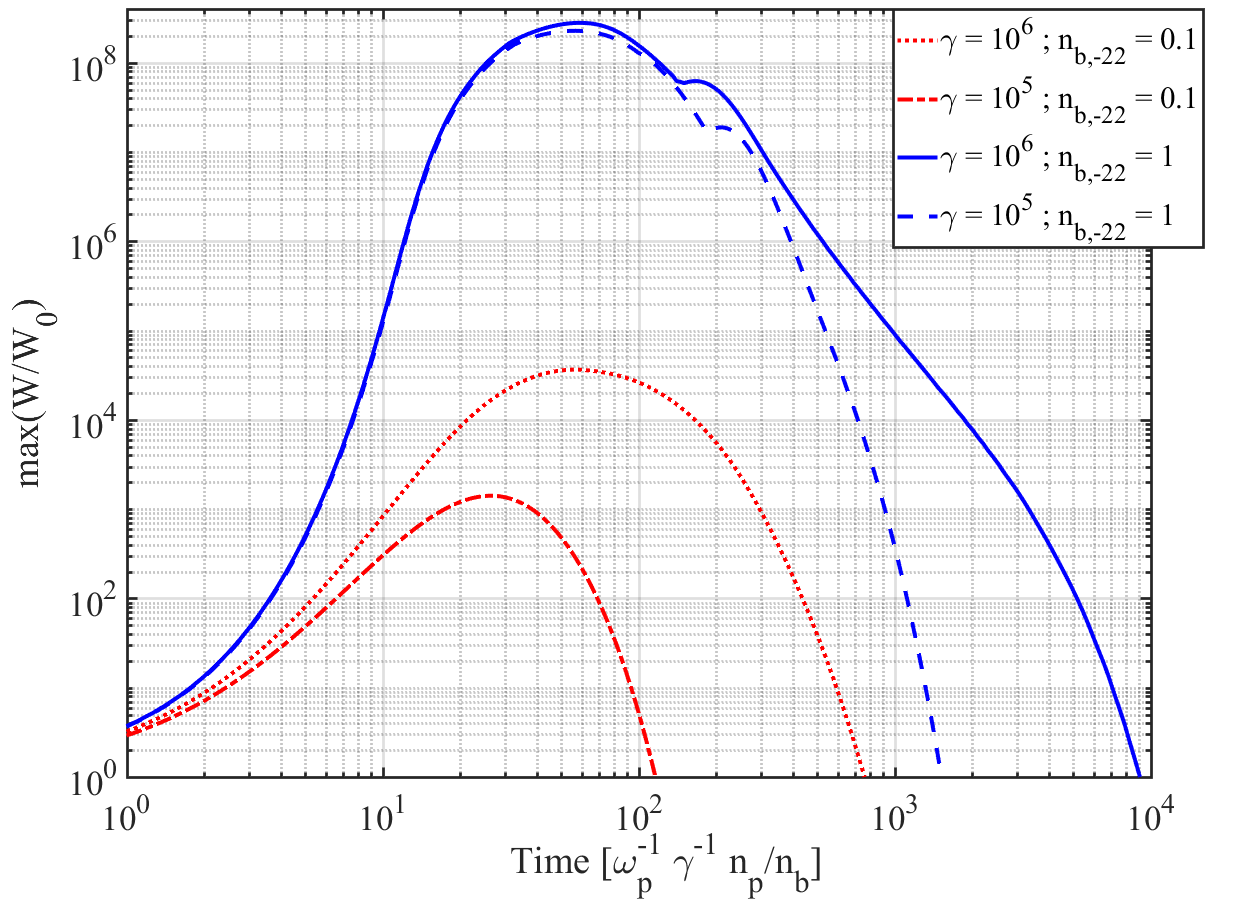}
    \caption{The maximum oscillation energy density $W_{osc}$, normalized by $W_0$ with respect to the entire resonant spectrum as a function of normalized time (\ref{eq:inst_time}) for $\gamma = 10^{6}, n_{b,-22} = 1$ (solid line), $\gamma = 10^{5}, n_{b,-22} = 1$ (dashed line), $\gamma = 10^{6}, n_{b,-22} = 0.1$ (dotted line), $\gamma = 10^{5}, n_{b,-22} = 0.1$ (dashed-dotted line). $L = 100Mpc$.}
   \label{fig:MAXW}
\end{figure}

It is interesting as well to examine the most energetic mode as a function of time and how it depends on the beam parameters. This is shown in figure \ref{fig:MAXW}. We have already seen in previous plots that the maximum value of $W_{\rm osc}$ initially grows 
but eventually begins to decrease. This 
occurs when inhomogeneity overcomes the linear growth, yet the competition between the two processes is dynamic and eventually the instability is completely suppressed. The maximum energy density with respect to the whole resonant spectrum strongly depends on the beam density, where a change in $n_b$ by a factor of 10 changes the possible growth of the energy density by orders of magnitude. 

\begin{figure}
    \centering
    \includegraphics[width=\columnwidth]{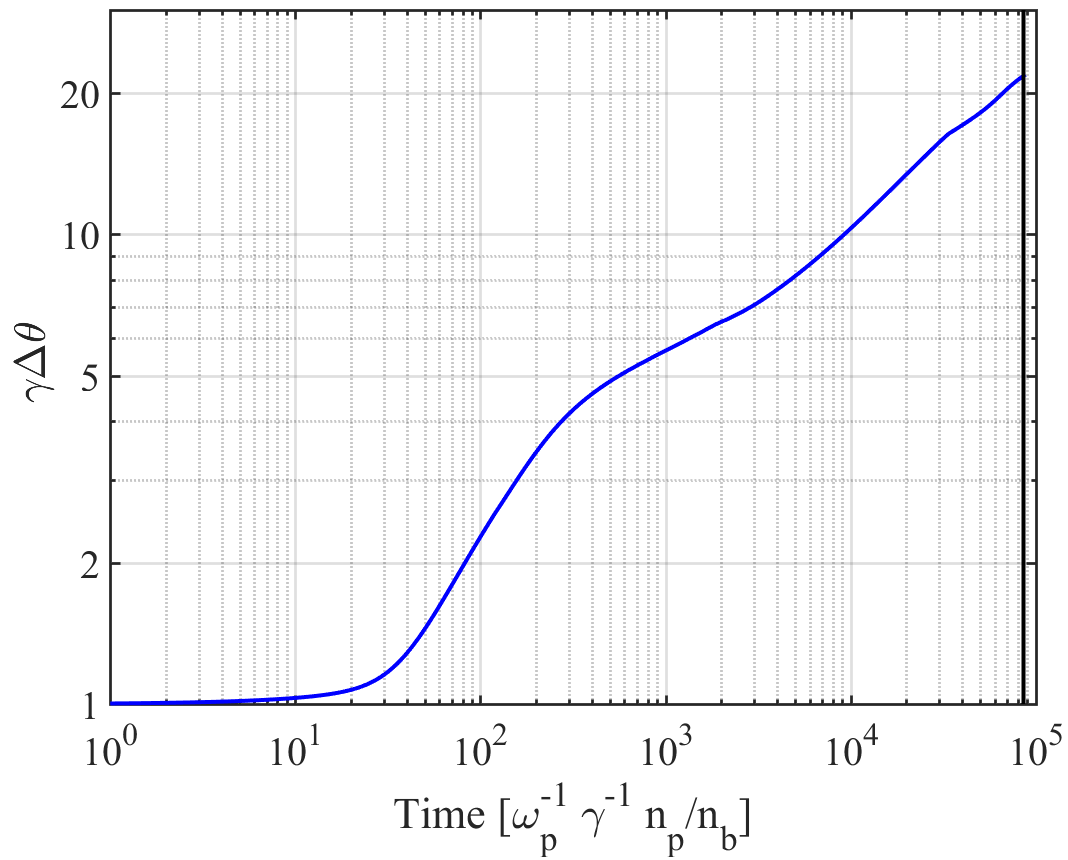}
    \caption{The evolution of the beam width when the collisional decay of the plasma waves is absent. Shown is the normalized angular spread, $\gamma\Delta\theta$, as a function of normalized time for $\gamma = 10^{6}, n_{b,-22} = 0.1$,  $L = 100$ Mpc. The typical inverse-Compton scattering time 
    is shown as a vertical black line.}
   \label{fig:DTHETA2}
\end{figure}

In the above simulations, the beam expands while the instability growth rate remains larger than the collisional decay time, after which the collisionless evolution ceases.  We also examine an artificial case where collisional decay is absent, i.e. $\nu_{\text{coll}} = 0$. For this case, waves will decay only if they exit the resonance range due to inhomogeneity. 
The waves propagating nearly perpendicularly to the direction of the density gradient  still grow so that the beam expands further out. The instability growth rate, and therefore the rate of evolution, decreases with the beam width. Eventually the beam expansion time becomes smaller than the Compton scattering time, after which the collisionless relaxation does not affect the beam any more: 
the beam will cool down and deposit its energy to the background light (\citealt{Broderick12,Sironi2014}).
Figure \ref{fig:DTHETA2} shows the expansion of the beam for this artificial case, for which $\gamma = 10^6$ and $n_{b,-22} = 0.1$. 
The simulations were stopped when the expansion time became equal to the characteristic Compton time.
One sees that the beams expands a few times more 
than in the case shown in figure \ref{fig:DTHETA_GAMMAS}. In any case, the beam evolution is determined by obliquely propagating waves, therefore the beam practically does not loose energy but just expands until the Compton scattering comes into the play. 

\subsection{Justification of the neglect of the energy evolution of the beam}
In the previous subsection, we studied the angular expansion of the beam. In addition to angular diffusion, the beam particles could exchange energy with the excited oscillations, which leads to a radial diffusion in momentum space. The energy exchange is governed by the small angles part of the spectrum of oscillations. In section\ 5.1 we have estimated the relative rate of the angular and energy diffusion and have shown that it depends both on the ratio between the diffusion coefficients and on the beam width, so that if the diffusion coefficients are of the same order of magnitude, then energy diffusion is negligible with respect to angular diffusion, see eq.\  (\ref{eq:diffusion_ratio}). According to our estimates, the diffusion coefficients are of the same order unless the plasma wave spectrum is dominated by small angle waves. Taking into account that growth of these waves is totally suppressed even by the slightest possible inhomogeneity, we retained only the angular terms in the momentum diffusion equation. Now we could justify our conjecture by directly calculating all the diffusion coefficients making use of the spectrum as obtained in simulations.  

\begin{figure}
    \centering
    \includegraphics[width=\columnwidth]{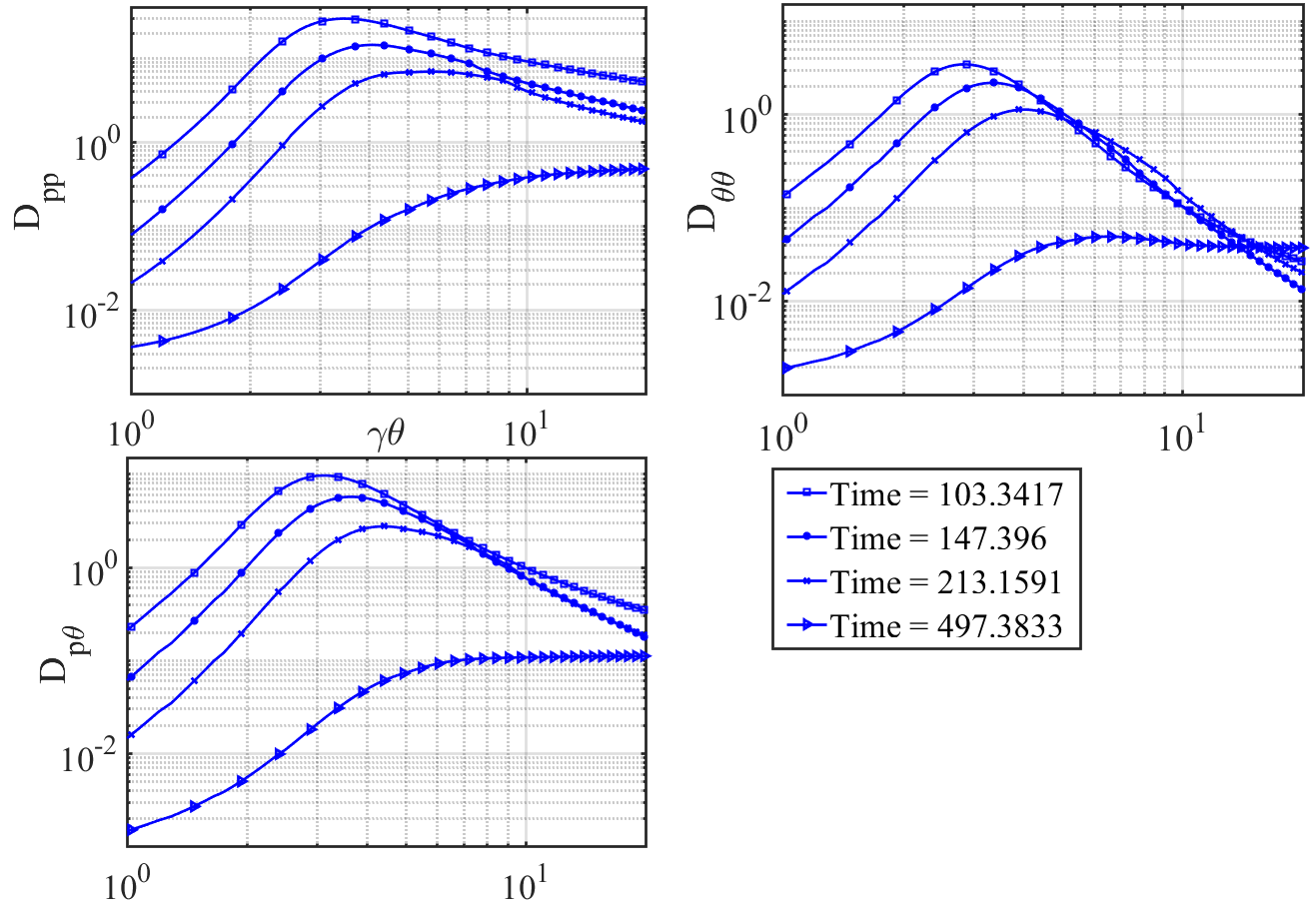}
    \caption{Angular dependence of the diffusion coefficients for different times: $t=103.34t_0$ (squares); $t=147.4t_0$ (circles); $t=213.16t_0$ (dashes);  $t=497.4t_0$ (triangles). 
    $n_{b,-22} = 1$ ,$ \gamma= 10^{6}$, $L = 100Mpc$.}
   \label{fig:Dtt_Dpp}
\end{figure}

\begin{figure}
    \centering
    \includegraphics[width=\columnwidth]{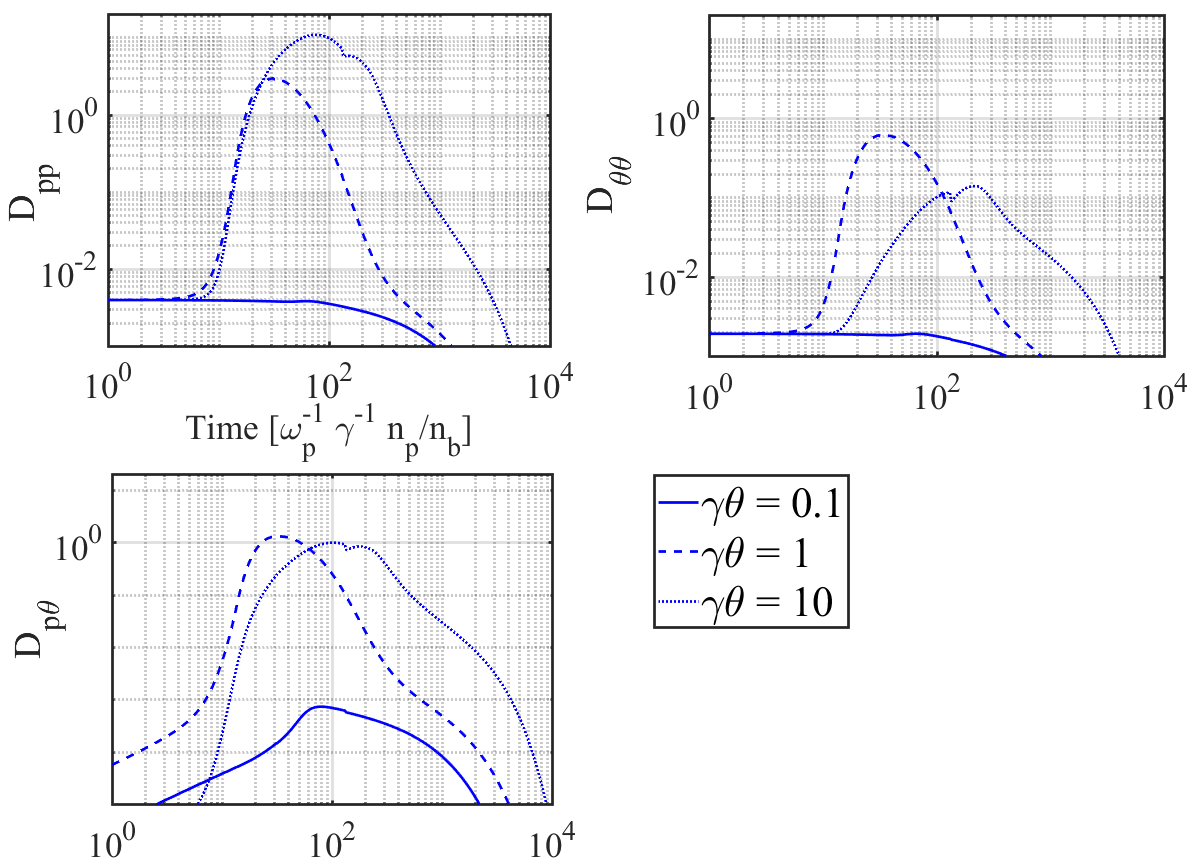}
    \caption{The time evolution of the diffusion coefficients at $\gamma\theta=0.1$ (solid line), $\gamma\theta=1$ (dashed) and $\gamma\theta=10$ (dotted).
     Beam parameters are the same as in the previous figure.}
   \label{fig:Dtt_Dpp_Dpt}
\end{figure}

Figure \ref{fig:Dtt_Dpp} shows the angular dependence of the diffusion coefficients for different times whereas figure \ref{fig:Dtt_Dpp_Dpt} shows the time evolution of the diffusion coefficients for different values of $\theta$. One sees that the diffusion coefficients remain of the same order all the time for all the relevant particle angles. This shows that the terms on the R.H.S. of equation (\ref{eq:Diff_eq}) are related as $1:\Delta\theta:\Delta\theta:(\Delta\theta)^2$, correspondingly. Taking into account that $\Delta\theta\sim 10^{-6} - 10^{-5}$, this justifies the neglect of the energy evolution of the beam. 

\section{Conclusions}
In this paper, we have examined the beam-plasma instability of Langmuir waves in the intergalactic medium. The source of the instability is the resonant energy exchange between a narrow, dilute, energetic beam of ultra-relativistic pairs traveling through the plasma and the Langmuir oscillations of the background plasma.
The beams of electrons and positrons which travel through the IGM are created by pair production processes of highly energetic gamma ray photons with the background radiation in the plasma. The evolution of the pair beams and the deposition of their energy to the IGM and back to the gamma ray flux is important to understand in order to explain the lack of any GeV halos around blazars \citep{Aharonian2006, Neronov2010, Broderick12}.

The crucial point is that 
the energetic evolution of the beam, whether it loses energy or gains transverse momentum, depends on the direction of the wave vector: if the interaction is dominated by waves propagating within the opening angle of the beam, $\theta_0\sim\Delta\theta$, then the beam loses a large fraction of its energy. However, when the interaction is dominated by oblique waves, the momentum loss is small and the beam simply expands. 

For realistic parameters of the IGM, the beam instability develops in
the kinetic regime. 
The kinetic growth rate dependence on the wave vector angle shows that 
the fastest growing waves are ones which travel at small angles ($\theta_0 \sim \Delta\theta$); their growth rate is a few times larger than that of oblique waves ($\theta_0 \sim 1$). This property of the kinetic instability explains the claim in the literature that the beam efficiently loses its energy \citep{Fainberg1970, Breizman71, Rudakov70} and therefore the instability is proposed as the principal mechanism of the beam evolution. However, the width of the wave-beam resonance for the nearly parallel waves is small $(\sim(\Delta\theta)^2)$, 
which means that in a very narrow beam, the growth of these waves is easily suppressed.  
We have shown that even for the slightest density gradient with a length scale as large as 100 Mpc, the quasi-parallel waves lose resonance 
at any 
reasonable parameters of intergalactic beams. 
Therefore the growth of the same waves which contribute to the decelerating mechanism is forbidden by the characteristics of the inhomogenous medium. 


We have solved the quasi-linear equations for the momentum diffusion of pair beams with a time evolution of the plasma oscillations, including the instability mechanism and the inhomogeneity term. We have found that the beam broadens in angle by about an order of magnitude but practically does not lose energy. 
This leads to the conclusion that the lack of GeV bumps in the blazar spectra cannot be attributed to the beam decay due to plasma instabilities. 

\section*{Acknowledgements}

This research was supported by the 
grant 2067/19 from the Israeli Science Foundation.

\section*{Data Availability}

The data underlying this article will be shared on reasonable
request to the corresponding author. 


\bibliographystyle{mnras}
\bibliography{example} 




\appendix

\section{Hydrodynamic instability}

The dispersion equation (3.1) for a mono-energetic beam is easily solved if $\omega_b\ll\omega_p$ (e.g., \S 61  in \citealt{Lifshitz_Pitaevskii81}). One can conveniently  write the
dispersion equation in the form
 \eqb{\label{eq:disp}}
1-\frac{\omega_p^2}{\omega^2}=\frac{\eta}{(\omega-\mathbf{k\cdot
v}_b)^2};
 \eqe
 where
 \eqb
\eta=\omega_b^2\frac{\gamma^2\sin\theta_0+\cos\theta_0}{\gamma^3}.
 \eqe
At $\omega_b\ll\omega_p$,  the rhs is non-negligible only if the
denominator is close to zero, therefore one can present the solution
in the form
 \eqb
\omega=\mathbf{k\cdot v}_b+\delta,
 \eqe
where $\delta\ll\omega$. Substituting this into (\ref{eq:disp}), one
finds
 \eqb
\delta=\pm\sqrt{\frac{\eta}{1-\frac{\omega^2_p}{k^2v^2_b}}}.
 \label{delta}\eqe
This solution represents  the well known beam mode. The wave is unstable at $kv<\omega_p$; the
growth rate is typically
 \eqb
 \Gamma_{1}\sim\eta^{1/2},
 \label{beam}\eqe
  but it increases when $kv_b$ approaches $\omega_p$. The
maximal growth rate is achieved at $kv\approx\omega\approx\omega_p$.
In this case, we deal with the plasma wave in resonance with the
beam. The solution is now presented as $\omega=\omega_p+\xi$, where
$\xi\ll \omega_p$. The standard procedure yields the growth rate of
the plasma wave
 \eqb
 \Gamma_{2}\sim(\eta\omega_p)^{1/3}.
 \label{res}\eqe
The ratio of the growth rate of the beam mode and of the plasma wave
is
 \eqb{\label{eq:ratio}}
\frac{\Gamma_{1}}{\Gamma_{2}}\sim\frac{\eta^{1/6}}{\omega_p^{1/3}}=\left(\frac{n_b}{n_p}\frac{\gamma^2\sin\theta_0+\cos\theta_0}{\gamma_b^3}\right)^{1/6}.
 \eqe
Due to the
power of $1/6$,  the growth rates of two modes could be roughly of
the same order for not very large Lorentz factors and moderate
density ratios.
Therefore in simulations, one can observe  the beam modes growing together with the
resonant plasma wave (e.g., \citealt{Bret_etal10}).  However, in the extragalactic
beams, $n_b/n_p\sim 10^{-15}-10^{-18}$, $\gamma_b\sim 10^6$;
therefore the
 above ratio (\ref{eq:ratio}) is as small as $10^{-4}-10^{-6}$.

The instability with such a small growth rate as that of the beam
modes could not develop in the IGM because of the collisional
damping. Regardless f this reason, one can readily see that the
beam modes do not in fact exist in the IGM. Namely, the condition
for these modes to exist is that the beam may be considered as
monoenergetic, so that all the beam particles are in resonance with
the wave,
 \eqb
\Gamma_1>k\Delta v.
 \label{condition}\eqe
In our case, $\Delta v\sim c\Delta\theta$, $k\sim\omega_p/c$. Even
for oblique waves, for which the growth rate is larger than for
parallel ones, we get
 \eqb
\frac{\Gamma_{1}}{\omega\Delta\theta}\sim\sqrt{\frac{n_b}{n_p\gamma_b}}\frac
1{\Delta\theta}\sim 10^{-6}-10^{-4},
 \eqe
 so that the condition (\ref{condition}) is violated by large
margin.

The above consideration shows that only resonant plasma waves could
be excited in the extragalactic beam-plasma systems therefore we consider only these waves in the paper.

\section{Kinetic growth rate for a Gaussian distribution}
Substituting into the general formula for the growth rate (\ref{eq:GW2}) the distribution function (\ref{Gauss}) and taking into account that $\theta \ll 1$, one gets:
\begin{equation}
\begin{gathered}
  \Gamma  = 8{\omega _p}\left( {\frac{1}{{\gamma \Delta {\theta ^4}}}} \right)\left( {\frac{{{n_b}}}{{{n_p}}}} \right){\left( {\frac{{{\omega _p}}}{{ck}}} \right)^3} \times  \hfill \\
  \int_{\theta _1^2}^{\theta _2^2} {d{\theta ^2}\frac{{\frac{{{\theta ^2}}}{2} + \left( {\frac{{2{{\cos }^2}{\theta _0}}}{{\theta _2^2 + \theta _1^2}} - 1 - \Delta {\theta ^2}} \right)}}{{\sqrt {\left( {{\theta ^2} - \theta _1^2} \right)\left( {\theta _2^2 - {\theta ^2}} \right)} }}{e^{ - \frac{{{\theta ^2}}}{{\Delta {\theta ^2}}}}}}  \hfill \\ 
\end{gathered} .
\end{equation}
Using the integral formulas of the modified Bessel functions of the first kind $I_{n} (x)$, 
\begin{equation}
{I_n}\left( x \right) = \frac{1}{\pi }\int_0^\pi  {{e^{x\cos \theta }}\cos \left( {n\theta } \right)d\theta },
\end{equation}
it is possible to evaluate the integrals
\begin{subequations}\label{eq:BesselIint}
\begin{flalign}
\begin{split}
&\int_{{x_1}}^{{x_2}} {\frac{{{e^{\frac{{2x}}{{\Delta {\theta ^2}}}}}}}{{\sqrt {\left( {x - {x_1}} \right)\left( {{x_2} - x} \right)} }}dx}  = \pi {e^{\frac{{{x_2} + {x_1}}}{{\Delta {\theta ^2}}}}}{I_0}\left( {\frac{{{x_2} - {x_1}}}{{\Delta {\theta ^2}}}} \right)\\
\end{split}, \\
\begin{split}
&\int_{{x_1}}^{{x_2}} {\frac{{x{e^{\frac{{2x}}{{\Delta {\theta ^2}}}}}}}{{\sqrt {\left( {x - {x_1}} \right)\left( {{x_2} - x} \right)} }}dx}  = \pi {e^{\frac{{{x_2} + {x_1}}}{{\Delta {\theta ^2}}}}} \times \\
&\left[ {\frac{{{x_2} - {x_1}}}{2}{I_1}\left( {\frac{{{x_2} - {x_1}}}{{\Delta {\theta ^2}}}} \right) + \frac{{{x_2} + {x_1}}}{2}{I_0}\left( {\frac{{{x_2} - {x_1}}}{{\Delta {\theta ^2}}}} \right)} \right].
\end{split}
\end{flalign}
\end{subequations}
Finally, we now get
\begin{equation}
\begin{gathered}
  \Gamma  = {\omega _p}\left( {\frac{{{n_b}}}{{{n_p}}}} \right)\frac{{2\pi }}{{\gamma \Delta {\theta ^2}}}{\cos ^3}{\theta _0}\exp \left( { - \frac{{\theta _1^2 + \theta _2^2}}{{2\Delta {\theta ^2}}}} \right) \times  \hfill \\
  \left[ {\left( {{\xi ^2}{{\left( {\frac{{\Delta \theta }}{{2{\theta _0}}}} \right)}^2} - 1} \right){I_0}\left( \xi  \right) - \frac{\xi }{2}{I_1}\left( \xi  \right)} \right], \hfill \\ 
\end{gathered},
\end{equation}
where $\xi  \equiv \left( {\theta _2^2 - \theta _1^2} \right)/2\Delta {\theta ^2}$.

\section{Resonance parameter}
{\label{SUBSEC:RES}}
The resonance parameter, $x$, is defined for each wave vector $(k,\theta_0)$ as
\begin{equation}{\label{eq:DefX1}}
x = \frac{\frac{\omega_{p}}{ck} - \cos\theta_0}{\sqrt{1-\frac{\omega_p}{ck}\cos\theta_0}}.
\end{equation}
First, we prove that $\left|x\right| \le \theta$ for resonant waves with particles at a polar angle $\theta$. We begin with the resonance condition
\begin{equation}
-1\le \frac{\omega_{p}/kc - \cos\theta \cos\theta_0 }{\sin\theta \sin\theta_0}\le1.
\end{equation}
Given that the polar angles are in the range $[0,\pi]$, one can get
\begin{equation}{\label{eq:RangeU}}
-\theta \sin {\theta _0} - \cos {\theta _0}\frac{{{\theta ^2}}}{2} \le \frac{{{\omega _p}}}{{kc}} - \cos {\theta _0} \le \theta \sin {\theta _0} - \cos {\theta _0}\frac{{{\theta ^2}}}{2}.
\end{equation}
After some arithmetics, we find the complete resonance range of $x$ as
\begin{equation}
 - \frac{{\theta \left( {1 + \frac{\theta }{{2\tan {\theta _0}}}} \right)}}{{\sqrt {1 + \frac{\theta }{{\tan {\theta _0}}} + \frac{{{\theta ^2}}}{{2{{\tan }^2}{\theta _0}}}} }} \le x \le \left\{ {\begin{array}{*{20}{c}}
   {\frac{{\theta \left( {1 - \frac{\theta }{{2\tan {\theta _0}}}} \right)}}{{\sqrt {1 - \frac{\theta }{{\tan {\theta _0}}} + \frac{{{\theta ^2}}}{{2{{\tan }^2}{\theta _0}}}} }}} & {{\theta _0} \ge  \theta }  \\ 
   {{\theta _0}/\sqrt 2 } & {{\theta _0} \le \theta }  \\ 
\end{array} } \right. .
\label{eq:resonance_range_x}\end{equation}

In order to express $k$ via $x$ and $\theta_0$, we write eq.\ (\ref{eq:DefX1}) as a
 quadratic equation:
\begin{equation}
{\left( {\frac{{{\omega _p}}}{{ck}}} \right)^2} + \left( {{x^2} - 2} \right)\cos {\theta _0}\frac{{{\omega _p}}}{{ck}} + {\cos ^2}{\theta _0} - {x^2} = 0.
\end{equation}
The solution is
\begin{equation}
\frac{{{\omega _p}}}{{ck}} =\left( {1 - \frac{{{x^2}}}{2}} \right)\cos {\theta _0} + \frac{x}{2}\sqrt {{x^2}{{\cos }^2}{\theta _0} + 4{{\sin }^2}{\theta _0}} ,
\end{equation}
where we kept only the solution which is consistent with (\ref{eq:DefX1}).
We divide the form of this solution to three different cases, or ranges of $x/ \theta_0$:
\begin{equation}
\begin{aligned}
&\frac{{{\omega _p}}}{{ck}} = \cos {\theta _0}\left(1 + x\tan {\theta _0}\right)  \quad ;\quad {\theta _0} \gg \left| x \right| ,\\
&\frac{{\omega_p}}{ck}  = 1 - \frac{\theta_0^2}{2} + \frac{x^2}{2} \left( \text{sgn}\left({x}\right) \sqrt{1+\frac{4\theta_0^2}{x^2}} -1 \right) \quad ;\quad {\theta _0} \sim \left| x \right| ,\\
&\frac{{{\omega _p}}}{{ck}} = 1 - {x^2}  \quad ;\quad {\theta _0} \ll \left| x \right| , x<0,
\end{aligned}
\end{equation}
where in the second and third cases we assumed that $\theta_0 \ll 1$.

\section{Kinetic growth rate in the general case}

Proceeding to a general distribution function, we first need to simplify the different terms in the growth rate formula. We begin by expressing the denominator in the growth rate formula (\ref{eq:GW2}) via $x$ and $\theta_0$:
\begin{equation}
\begin{aligned}
&\left( {\cos {\theta _2} - \cos \theta } \right)\left( {\cos \theta  - \cos {\theta _1}} \right) = \\
&{\sin ^4}{\theta _0}\left[ \begin{array}{l}
1 - \frac{{2x}}{{\tan {\theta _0}}} + \frac{{{x^2}}}{{{{\tan }^2}{\theta _0}}} - {x^2}{\left( {1 - \frac{x}{{2\tan {\theta _0}}}} \right)^2} - \\
{\left\{ {1 - \frac{x}{{\tan {\theta _0}}} + \frac{{{x^2}}}{{2{{\tan }^2}{\theta _0}}} - \frac{{{\theta ^2}}}{{2{{\sin }^2}{\theta _0}}}} \right\}^2}
\end{array} \right].
\end{aligned}
\end{equation}
We next consider the formula in three different regimes of $x/\theta_0$.
\begin{enumerate}
\item{$\left|x\right| \ll \theta_0$}
\begin{equation}
\frac{\omega_{p}}{ck} = \cos\theta_0 + x\sin\theta_0,
\end{equation}
\begin{equation}
\begin{aligned}
\left( {\cos \theta  - \cos {\theta _1}} \right)\left( {\cos {\theta _2} - \cos \theta } \right) =& \\ {\sin ^2}{\theta _0} - {\cos ^2}\theta  - {\left( {\frac{{{\omega _p}}}{{ck}}} \right)^2} + &\frac{{2{\omega _p}}}{{ck}}\cos {\theta _0}\cos \theta .
\end{aligned}
\end{equation}
After some arithmetics, one gets:
\begin{equation}
\left( {\cos \theta  - \cos {\theta _1}} \right)\left( {\cos {\theta _2} - \cos \theta } \right) = {\sin ^2}{\theta _0}\left( {{\theta ^2} - {x^2}} \right).
\end{equation}
The nominator can be rewritten with a change of variables:
\begin{equation}
\begin{array}{l}
\left[ {\frac{1}{\theta }\frac{{\partial g}}{{\partial \theta }}\left( {\cos \theta  - \frac{{ck}}{{{\omega _p}}}\cos {\theta _0}} \right) - 2g} \right]\theta d\theta  = \\
\left[ {\frac{{\partial g}}{{\partial \left( {{\theta ^2}/{x^2}} \right)}}\left( {x\tan {\theta _0} - {x^2}\left( {{{\tan }^2}{\theta _0} + \frac{{{\theta ^2}}}{{2{x^2}}}} \right)} \right) - g{x^2}} \right]d\left( {\frac{{{\theta ^2}}}{{{x^2}}}} \right).
\end{array}
\end{equation}
Combining all these together, the growth rate is written as an integral over a single variable $z = \theta^2/x^2$:
\begin{equation}{\label{eq:GW3}}
\begin{aligned}
&\Gamma  = \frac{{{\omega _p}}}{2}\left( {\frac{{mc}}{{{n_b}}}} \right)\left( {\frac{{{n_b}}}{{{n_p}}}} \right){\cos ^2}{\theta _0}\left( {1 + 3x\tan {\theta _0}} \right) \times \\
&\int_1^\infty  {\frac{{\frac{{\partial g}}{{\partial z}}\left[ {{\mathop{\rm sgn}} (x) - \left( {\frac{{\left| x \right|}}{{\tan {\theta _0}}}} \right)\left( {{{\tan }^2}{\theta _0} + \frac{z}{2}} \right)} \right] - g\left( {\frac{{\left| x \right|}}{{\tan {\theta _0}}}} \right)}}{{\sqrt {z - 1} }}dz} \\ &\approx 
 \frac{{{\omega _p}}}{2}\left( {\frac{{mc}}{{{n_b}}}} \right)\left( {\frac{{{n_b}}}{{{n_p}}}} \right){\rm{sgn}}(x){\cos ^2}{\theta _0}\int_1^\infty  {\frac{{\frac{{\partial g}}{{\partial z}}}}{{\sqrt {z - 1} }}dz} .
\end{aligned}
\end{equation}
\item{$\left|x\right | \sim \theta_0$}
\newline
First we evaluate the denominator terms:
\begin{equation}
\begin{gathered}
  \cos {\theta _{1,2}} - \cos \theta  = \frac{{{\omega _p}}}{{ck}}\cos {\theta _0} \pm \sin {\theta _0}\sqrt {1 - {{\left( {\frac{{{\omega _p}}}{{ck}}} \right)}^2}}  - \cos \theta  =  \hfill \\
  \frac{{{\theta ^2}}}{2} - \theta _0^2 + \frac{{{x^2}}}{2}\left[ {{\text{sgn}}(x)\sqrt {1 + \frac{{4\theta _0^2}}{{{x^2}}}}  - 1} \right] \pm  \hfill \\
  {\theta _0}\left| x \right|\sqrt {\frac{{\theta _0^2}}{{{x^2}}} + 1 - {\text{sgn}}(x)\sqrt {1 + \frac{{4\theta _0^2}}{{{x^2}}}} }  =  \hfill \\
   = \frac{{{x^2}}}{2}\left\{ \begin{gathered}
  \frac{{{\theta ^2}}}{{{x^2}}} - \frac{{2\theta _0^2}}{{{x^2}}} + \left[ {{\text{sgn}}(x)\sqrt {1 + \frac{{4\theta _0^2}}{{{x^2}}}}  - 1} \right] \pm  \hfill \\
  \frac{{2{\theta _0}}}{{\left| x \right|}}\sqrt {\frac{{\theta _0^2}}{{{x^2}}} + 1 - {\text{sgn}}(x)\sqrt {1 + \frac{{4\theta _0^2}}{{{x^2}}}} }  \hfill \\ 
\end{gathered}  \right\}. \hfill \\ 
\end{gathered}.
\end{equation}
Then the denominator term reduces to
\begin{equation}
\begin{gathered}
  \left( {\cos {\theta _2} - \cos \theta } \right)\left( {\cos \theta  - \cos {\theta _1}} \right) =  \hfill \\
  \frac{{{x^4}}}{4}\left[ \begin{gathered}
  4\mu \left[ {\mu  + 1 - {\text{sgn}}(x)\sqrt {1 + 4\mu } } \right] -  \hfill \\
  {\left( {z - 2\mu  + \left[ {{\text{sgn}}(x)\sqrt {1 + 4\mu }  - 1} \right]} \right)^2} \hfill \\ 
\end{gathered}  \right] \hfill \\
   = \frac{{{x^4}}}{2}\left[ {\left( {2\mu  - \rho } \right)\left( {z - 1} \right) - \frac{{{z^2}}}{2}} \right], \hfill \\ 
\end{gathered} ,
\end{equation}
where we have defined $z = \theta^2/x^2$, $\mu = \theta_0^2/x^2$ and $\rho =  {\mathop{\rm sgn}} (x)\sqrt {1 + 4\mu }  - 1$.

The roots of the denominator, which are also the integral boundaries, are given by solving the proper quadratic equation for $z$: 
\begin{equation}{\label{eq:Zroots}}
z_{1,2} = 2\mu  -\rho \pm 2  \sqrt {\mu\left(\mu  - \rho\right)} .
\end{equation}
The nominator is evaluated using the definitions of $z,\mu$ and $\rho$ above. Given the expression
\begin{equation}
\frac{{{\omega _p}}}{{ck}} = 1 - \frac{{{x^2}}}{2}\left( {\mu  - \rho } \right),
\end{equation}
the nominator can be written as
\begin{equation}
\left[ {\frac{1}{\theta }\frac{{\partial g}}{{\partial \theta }}\left( {\cos \theta  - \frac{{ck}}{{{\omega _p}}}\cos {\theta _0}} \right) - 2g} \right]\theta d\theta  = \left[ {\frac{1}{2}\frac{{\partial g}}{{\partial z}}\left( {\rho  - z} \right) - g} \right]{x^2}dz.
\end{equation}
Combining all these together, the expression for the growth rate is
\begin{equation}{\label{eq:GW4}}
\Gamma  = {\omega _p}\left( {\frac{{mc}}{{{n_b}}}} \right)\left( {\frac{{{n_b}}}{{{n_p}}}} \right)\int_{{z_1}}^{{z_2}} {\frac{{\frac{1}{2}\frac{{\partial g}}{{\partial z}}\left( {\rho  - z} \right) - g}}{{\sqrt {\left( {z - {z_1}} \right)\left( {{z_2} - z} \right)} }}dz} .
\end{equation}
\item{$\left|x\right | \gg \theta_0$ , $ x < 0$}
\begin{equation}
\begin{aligned}
&\frac{{{\omega _p}}}{{ck}} \approx 1 - {x^2} - \frac{{3\theta _0^2}}{2},\\
&\sqrt {\left( {\cos {\theta _2} - \cos \theta } \right)\left( {\cos \theta  - \cos {\theta _1}} \right)}  = \frac{{{x^2}}}{2}\sqrt {\left( {z - 2} \right)\left( {2 + 8\mu  - z} \right)} ,\\
&\left[ {\frac{1}{\theta }\frac{{\partial g}}{{\partial \theta }}\left( {\cos \theta  - \frac{{ck}}{{{\omega _p}}}\cos {\theta _0}} \right) - 2g} \right]\theta d\theta  =  - \left[ {\frac{{\partial g}}{{\partial z}}\left( {\frac{z}{2} + 1} \right) + g} \right]{x^2}dz.\\
\end{aligned}
\end{equation}
All together, we get
\begin{equation}
    \Gamma  = \pi {\omega _p}\left( {\frac{{mc}}{{{n_b}}}} \right)\left( {\frac{{{n_b}}}{{{n_p}}}} \right){\left[ {\left( { -2 \frac{{\partial g}}{{\partial z}}} \right)- g} \right]_{z = 2}}.
\end{equation}
\end{enumerate}
The evaluation of the growth rates in section \ref{sec:KIN} and in the numerical simulation is done by using these three cases, given that the formulas match each other at intermediate ranges of $x/\theta_0$.

\section{Diffusion coefficients}
\begin{figure}
    \centering
     \includegraphics[width=\columnwidth]{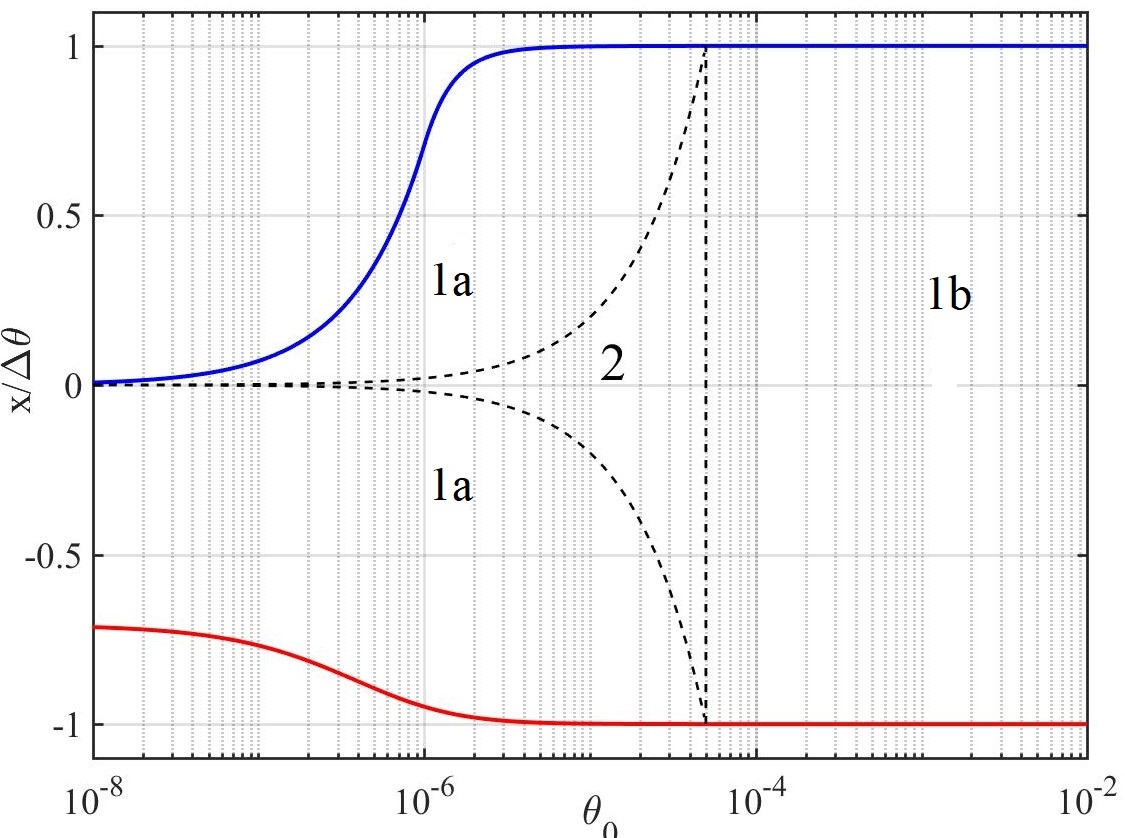}
    \caption{Integration space for the diffusion coefficient at $\theta = \Delta\theta = 10^{-6}$. The solid lines are the curves for the upper and lower bounds of $x$ as a function of $\theta_0$, respectively. The dashed lines represent the boundaries of the two integration subspaces, 
    denoted by $1$ and $2$ in the figure, respectively. The subdivision of subspace '1' is denoted by '1a' and '1b' and their boundaries are also denoted by dashed lines. The vertical dashed line represents the line $\theta_0 = 50\theta$.}
    \label{fig:XBOUNDS}
\end{figure}

{\label{subSec:DiffCoeff}}
The evaluation of diffusion coefficients (\ref{eq:DIFF_COEFF}) as a function of the beam angle $\theta$ requires an integration over the entire range of waves resonant with a pair-beam particle with polar angle $\theta$. 
After the $\phi_0$ integration, using the delta function, the diffusion coefficient takes the form
\begin{equation}
{D_\mu } = \pi {e^2}\int {\frac{{W\left( {{\mathbf{k}},t} \right)}}{{{\theta ^\mu }}}\frac{{{{\left( {\frac{{{\omega _p}}}{{kc}} - \cos {\theta _0} - \frac{{{\theta ^2}\cos {\theta _0}}}{2}} \right)}^\mu }{k^2}dkd\cos {\theta _0}}}{{kc\sqrt {\left[ {\cos {\theta _0} - {{\left( {\cos {\theta _0}} \right)}_1}} \right]\left[ {{{\left( {\cos {\theta _0}} \right)}_2} - \cos {\theta _0}} \right]} }}}  .
\end{equation}
This expression is too cumbersome to evaluate, even numerically. One has to simplify the expressions in the integral by dividing the integration space to two non-overlapping sub-spaces, based on the magnitude of $x/\theta_0$ and $\theta_0/\theta$. The complete diffusion coefficient integral is then presented as a sum of two expressions,
\begin{equation}{\label{eq:DIFF_SEP}}
    D_{\mu} = D_{1,\mu} + D_{2,\mu},
\end{equation}
where subscript '1' corresponds to large values of $\theta_0/x$ or $\theta/\theta_0$ and '2' to the $(x,\theta_0)$ space complementary.
The first subspace, '1', contains all points $\left(x,\theta_0\right)$ for which $\theta_0$ is much larger than $\theta$ or than $\left|x\right|$. The second subspace contains all points $\left(x,\theta_0\right)$ for which $\theta_0$ is not much larger than $\left|x\right|$ and $\theta$.
Subspace '1' is separated further into two smaller non-overlapping sub-spaces '1a' and '1b' as follows:
Subspace '1a' ,$V_{1a}$, contains all points $(x,\theta_0)$ such that $50 \left|x\right| < \theta_0$ and $\theta_0 \le 50 \theta$.
Subspace '1b' ,$V_{1b}$, contains all points $(x,\theta_0)$ such that $\theta_0 > 50 \theta$.
We have found that the arbitrary choice of the large factor of 50 is suitable for the approximations made in the simplifications of the algebraic expressions.
Subspace '2' contains all the rest. The sub-division of subspace '1' is mainly for convenience, while the integration formula is the same.
An example for the integration space $\left(x,\theta_0\right)$ for the diffusion coefficients is described in figure \ref{fig:XBOUNDS}. The solid curves correspond to the boundaries of the resonance parameter, $x$, as determined by formula (\ref{eq:resonance_range_x}). 
The complete integration space is then the area between these two curves (ranging over all values of $\theta_0$).
The dashed lines correspond to the separation between the two sub-spaces, where the numbers '1' and '2' denote the subspace.

We now proceed to algebraic development of the diffusion integration formulas for each subspace separately.
In the subspace 1, $\theta, \left|x\right | \ll \theta_0$. Expanding in small parameters to the first non-vanishing order, one gets, after some work,
\begin{equation}{\label{eq:DiffCo_1}}
D_{1,\mu } = \frac{{\pi {e^2}\omega _p^2}}{{2{c^3}}}\int_{{V_1}\left( \theta  \right)} {{{\left( {\frac{{{{\tan }^2}{\theta _0}}}{{1 + {{\tan }^2}{\theta _0}}}} \right)}^{\mu /2}}\frac{{{z^\mu }}}{{\sqrt {1 - {z^2}} }}W\left( {{\mathbf{k}},t} \right)dzd{{\tan }^2}{\theta _0}} , 
\end{equation}
where $z=x/\theta$.

In the subspace 2, we  define  new variables 
\begin{equation}{\label{eq:Defy}}
y = \frac{{x\sqrt {{x^2} + 4\theta _0^2}  - {x^2}}}{{{\theta ^2}}} + 1;\qquad w=\frac{\theta_0}{\theta}.
\end{equation}
Then the diffusion coefficient can be written, to zeroth order in $\theta$, as
\begin{equation}{\label{eq:DiffCo_2}}
D_{2,\mu }^0 = \frac{{2\pi {e^2}\omega _p^2}}{{{c^3}}}{\left( {\frac{\theta }{2}} \right)^{\mu  + 2}}\int_{{V_2}\left( \theta  \right)} {W\left( {{\mathbf{k}},t} \right)\frac{{{{\left( {y - 2} \right)}^\mu }}}{{\sqrt {4{w^2} - {y^2}} }}d{w^2}dy} .
\end{equation}
\bsp	
\label{lastpage}
\end{document}